\documentclass[longauth]{aa} 
\usepackage{graphicx}
\usepackage{txfonts}
\usepackage{orcidlink}

\usepackage{newtxtext,newtxmath}
\usepackage[T1]{fontenc}
\usepackage{ae,aecompl}
\usepackage{multirow}
\usepackage{tabularx}
\usepackage{graphicx}
\usepackage{array}
\usepackage{supertabular}
\usepackage{longtable}
\usepackage[]{url}
\usepackage{graphicx}
\usepackage{comment}
\usepackage{booktabs}
\usepackage{gensymb}
\usepackage{xcolor,color}
\usepackage[caption = false]{subfig}
\usepackage{ulem}
\usepackage[cm]{aeguill}

\usepackage{float}
\usepackage{rotating}
\usepackage{amsmath}
\usepackage{threeparttable}
\usepackage{hyperref}

\usepackage{xfp}
\usepackage{calc}
\newlength{\mynumberwidth}
\newlength{\myminuswidth}
\DeclareUnicodeCharacter{0301}{é}

\definecolor{darkgreen}{RGB}{0,100,0}

\usepackage{tabularx}
\usepackage{booktabs}

\newcommand{\afterglowpy}{\textsc{afterglowpy}}

\newcommand{\red}[1]{\textcolor{black}{#1}}
\DeclareRobustCommand{\VAN}[3]{#2}
\let\VANthebibliography\thebibliography
\def\thebibliography{\DeclareRobustCommand{\VAN}[3]{##3}\VANthebibliography}

\hypersetup{
 colorlinks=true,
 allcolors=blue
}

\begin{document}

\title{GRB 241030A: a bright afterglow challenging forward shock emission}
%\titlerunning{GRB 241030A: a bright afterglow challenging the standard fireball model}

\author{
  % Science co-lead
  J.-G.~Ducoin\,\orcidlink{0009-0008-7341-4825}\inst{\ref{aix_marseille_cppm_iphu}}
  \and
  % Science  co-lead
  C.~Pellouin\,\orcidlink{0000-0001-9489-783X}\inst{\ref{tau_physics}}
  \and
  % AbAO - observer
  V.~Aivazyan\inst{\ref{georgian_national_astro_obs},\ref{sjsu_georgia}}
  \and
  % PWT
  D.~Akl\,\orcidlink{0009-0006-4358-9929}\inst{\ref{univ_cote_azur},\ref{nYu_abudhabi_cass},\ref{nYu_abudhabi}}
  \and
  % COLIBRI
  F.~Alvarez\,\orcidlink{0009-0009-5612-3759}\inst{\ref{unam_astronomy}}
  \and
  % GRANDMA Core Team  - KNC
  C.~Andrade\,\orcidlink{0009-0004-9687-3275}\inst{\ref{umn_physics}}
  \and
  % COLIBRI
  C.~Angulo\,\orcidlink{0009-0002-6667-3294}\inst{\ref{unam_astronomy}}
  \and
  % PWT, GRANDMA core-team
  S.~Antier\,\orcidlink{0000-0002-7686-3334}\inst{\ref{univ_cote_azur},\ref{ijclab_orsay}}
  \and
  % COLIBRI
  J.-L.~Atteia\,\orcidlink{0000-0001-7346-5114}\inst{\ref{irap_toulouse}}
  \and
  % COLIBRI PI
  S.~Basa\,\orcidlink{0000-0002-4291-333X}\inst{\ref{aix_marseille_lam}}
  \and
  % COLIBRI
  R.L.~Becerra,\orcidlink{0000-0002-0216-3415}\inst{\ref{unam_astronomy}}
  \and
  % GRANDMA HAO
  Z.~Benkhaldoun\inst{\ref{astro_sharjah},\ref{oukaimeden_obs}}
  \and
  % LAT
  E.~Bissaldi\,\orcidlink{0000-0001-9935-8106}\inst{\ref{poliba_physics},\ref{infn_bari}}
  \and
  % Swift
  A.~Breeveld\,\orcidlink{0000-0002-0001-7270}\inst{\ref{mssl_ucl}}
  \and
  % GRANDMA Data Analysis
  E.~de.~Bruin\inst{\ref{umn_physics}}
  \and
  % GRANDMA, GBM analysis
  E.~Burns\inst{\ref{lsu_physics}}
  \and
  % COLIBRI
  N.R.~Butler\inst{\ref{asu_earth_space}}
  \and
  % GRANDMA Core Team
  M.W.~Coughlin\,\orcidlink{0000-0002-8262-2924}\inst{\ref{umn_physics}}
  \and
  % GRANDMA Review
  F.~Daigne\inst{\ref{iap_paris}}
  \and
  % NMMA Modelling
  T.~Dietrich\inst{\ref{potsdam_astro},\ref{aei_potsdam}}
  \and
  % COLIBRI observer - GRANDMA
  D.~Dornic\,\orcidlink{0000-0001-5729-1468}\inst{\ref{aix_marseille_cppm_iphu}}
  \and
  % Skyportal
  C.~Douzet\,\orcidlink{0009-0000-4541-2074}\inst{\ref{ijclab_orsay}}
  \and
  % Skyportal
  T.~du~Laz\inst{\ref{caltech_astro}}
  \and
  % GRANDMA Core Team - DAG - Data analysist on KAO
  P.-A.~Duverne\,\orcidlink{0000-0002-3906-0997}\inst{\ref{apc_paris}}
  \and
  % KNC
  H.B.~Eggenstein\,\orcidlink{0000-0001-5296-7035}\inst{\ref{knc_paderborn}}
  \and
  % KAO - Observer
  E.~Elhosseiny\,\orcidlink{0000-0002-9751-8089}\inst{\ref{nriag_egypt}}
  \and
  % TNOT Observer
  A.~Esamdin\inst{\ref{xao_china},\ref{ucas_astronomy}}
  \and
  % Swift
  P.A.~Evans\,\orcidlink{0000-0002-8465-3353}\inst{\ref{leicester_astro}}
  \and
  % GRANDMA GTC
  J.~F.~Agüí~Fernández\,\orcidlink{0000-0001-6991-7616}\inst{\ref{calar_alto}}
  \and
  % Swift
  M.~Ferro\,\orcidlink{0009-0007-5708-7978}\inst{\ref{brera_obs}}
  \and
  % COLIBRI
  F.~Fortin\inst{\ref{irap_toulouse}}
  \and
  % KNC observer
  M.~Freeberg\inst{\ref{colfax_usa}}
  \and
  % COLIBRI
  L.~Garc\'ia-Garc\'ia\,\orcidlink{0000-0001-5125-1043}\inst{\ref{unam_astronomy}}
  \and
  % COLIBRI
  R.~Gill\,\orcidlink{0000-0003-0516-2968}\inst{\ref{iraf_unam}}
  \and
  % COLIBRI
  N.~Globus\,\orcidlink{0000-0001-9011-0737}\inst{\ref{unam_ensenada}}
  \and
  % GRANDMA Core Team - co-Coordinator - Reviewer
  N.~Guessoum\,\orcidlink{0000-0003-1585-8205}\inst{\ref{aus_physics}}\thanks{Corresponding author: \email{nguessoum@aus.edu}}
  \and
  % KAO - Observer
  G.M.~Hamed\,\orcidlink{0000-0001-6009-1897}\inst{\ref{nriag_egypt}}
  \and
  % GRANDMA Core Team - Co-PI
  P.~Hello\inst{\ref{ijclab_orsay}}
  \and
  % LAT
  A.~Holzmann~Airasca\,\orcidlink{0009-0007-8169-4719}\inst{\ref{univ_trento},\ref{infn_bari}}
  \and
  % EP
  D.F.~Hu\inst{\ref{pmo_china},\ref{ustc_astronomy}}
  \and
  % GRANDMA shifter
  T.~Hussenot-Desenonges\,\orcidlink{0009-0009-2434-432X}\inst{\ref{ijclab_orsay}}
  \and
  % AbAO - Observer
  R.~Inasaridze\,\orcidlink{0000-0002-6653-0915}\inst{\ref{georgian_national_astro_obs},\ref{sjsu_georgia}}
  \and
  % TNOT Observer
  A.~Iskandar\,\orcidlink{0009-0003-9229-9942}\inst{\ref{xao_china},\ref{ucas_astronomy}}
  \and
  % EP
  S.Q.~Jiang\inst{\ref{naoc_china},\ref{ucas_astronomy_2}}
  \and
  % EP
  C.C.~Jin\inst{\ref{naoc_china},\ref{ucas_astronomy_2},\ref{bnu_astro}}
  \and
  % GRANDMA HAO
  A.~Kaeouach\inst{\ref{oukaimeden_obs}}
  \and
  % GRANDMA Core Team - FRAM - DA of the Data
  S.~Karpov\,\orcidlink{0000-0003-0035-651X}\inst{\ref{fzu_praha}}
  \and
  % Swift
  N.~J.~Klingler\,\orcidlink{0000-0002-7465-0941}\inst{\ref{nasa_goddard},\ref{umbc_space},\ref{crest_nasa}}
  \and
  % TAROT -PI
  A.~Klotz\,\orcidlink{0000-0003-0106-4148}\inst{\ref{irap_toulouse}}
  \and
  % AbAO - PI Observer Operations DA
  N.~Kochiashvili\,\orcidlink{0000-0001-5249-4354}\inst{\ref{georgian_national_astro_obs}}
  \and
  % GRANDMA NMMA
  H.~Koehn\,\orcidlink{0009-0001-5350-7468}\inst{\ref{potsdam_astro}}
  \and
  % KNC
  R.~Kneip\,\orcidlink{0009-0009-1964-677X}\inst{\ref{knc_contern}}
  \and
  % AbAO - Observer
  T.~Kvernadze\inst{\ref{georgian_national_astro_obs}}
  \and
  % Skyportal
  A.~Le~Calloch\orcidlink{0009-0009-7000-8343}\inst{\ref{umn_physics}}
  \and
  % COLIBRI
  W.H.~Lee\inst{\ref{unam_astronomy}}
  \and
  % KNC
  A.~Lekic\,\orcidlink{0009-0007-2048-4865}\inst{\ref{knc_ipsa}}
  \and
  % EP
  Y. F. Liang\,\orcidlink{0009-0005-0170-192X}\inst{\ref{pmo_china},\ref{ustc_astronomy}}\thanks{Corresponding author: \email{yfliang@pmo.ac.cn}}
  \and
  % TAROT - oper
  C.~Limonta\inst{\ref{univ_cote_azur}}
  \and
  % TNOT Observer
  J.~Liu\inst{\ref{tsinghua_physics}}
  \and
  % COLIBRI
  K.~Ocelotl.~C.~L\'opez\,\orcidlink{0000-0002-9322-6900}\inst{\ref{unam_astronomy}}
  \and
  % COLIBRI
  D.~L\'opez-C\'amara\,\orcidlink{0000-0001-9512-4177}\inst{\ref{icn_unam}}
  \and
  % KAO - Observer
  R.H.~Mabrouk\,\orcidlink{0009-0003-7202-4159}\inst{\ref{nriag_egypt}}
  \and
  % COLIBRI/GRANDMA - Data analysis
  F.~Magnani\inst{\ref{aix_marseille_cppm_iphu}}
  \and
  % GRANDMA interpretation - PWT
  J.~Mao\,\orcidlink{0000-0002-7077-7195}\inst{\ref{ynao_china},\ref{cams_china},\ref{klo_china}}
  \and
  % FRAM Auger
  M.~Mašek\,\orcidlink{0000-0002-0967-0006}\inst{\ref{fzu_praha}}
  \and
  % COLIBRI -
  E.~Moreno~M\'endez\,\orcidlink{0000-0002-5411-9352}\inst{\ref{fac_ciencias_unam}}
  \and
  % NAO - Observer
  B.M.~Mihov\,\orcidlink{0000-0002-1567-9904}\inst{\ref{inao_bulgaria}}
  \and
  % GRANDMA - KAO
  M.~Molham\,\orcidlink{0000-0002-3072-8671}\inst{\ref{nriag_egypt}}
  \and
  % GRANDMA - TRT
  K.~Noysena\inst{\ref{narit_thailand}}
  \and
  % GRANDMA - KNC
  M.~Odeh\,\orcidlink{0000-0002-8986-6681}\inst{\ref{knc_iac}}
  \and
  % LAT
  N.~Omodei\,\orcidlink{0000-0002-5448-7577}\inst{\ref{stanford_kipac}}
  \and
  % TNOT
  H.~Peng\inst{\ref{tsinghua_physics}}
  \and
  % COLIBRI
  M.~Pereyra
  \orcidlink{0000-0001-6148-6532}\inst{\ref{unam_astronomy}}
  \and
  % GRANDMA - GRB Chair - Coordinator
  M.~Pillas\,\orcidlink{0000-0003-3224-2146}\inst{\ref{iphc_strasbourg},\ref{iap_paris}}\thanks{Corresponding author: \email{marion.pillas@iap.fr}}
  \and
  % LAT
  R.~Pillera\,\orcidlink{0000-0003-3808-963X}\inst{\ref{infn_bari}}
  \and
  % GRANDMA Core Team
  T.~Pradier\,\orcidlink{0000-0001-5501-0060}\inst{\ref{iphc_strasbourg}}
  \and
  % GRANDMA operations
  Y.~Rajabov\inst{\ref{ulugh_beg_inst}}
  \and
  % PWT - GRANDMA - Colibri
  N.A.~Rakotondrainibe\,\orcidlink{0009-0004-0263-7766}\inst{\ref{aix_marseille_lam}}
  \and
  % COLIBRI Team
  B.~Schneider\,\orcidlink{0000-0003-4876-7756}\inst{\ref{aix_marseille_lam}}
  \and
  % KNC
  M.~Serrau\,\orcidlink{0009-0003-5793-4293}\inst{\ref{knc_saf}}
  \and
  % NAO - Observer - GRANDMA
  L.~Slavcheva-Mihova\,\orcidlink{0000-0002-1582-4913}\inst{\ref{inao_bulgaria}}
  \and
  % GRANDMA
  O.~Sokoliuk\,\orcidlink{0000-0003-4503-7272}\inst{\ref{kyiv_astro_obs},\ref{aberdeen_physics}}
  \and
  % EP
  H.~Sun\,\orcidlink{0000-0002-9615-1481}\inst{\ref{naoc_china}}
  \and
  % KAO - PI
  A.~Takey\,\orcidlink{0000-0003-1423-5516}\inst{\ref{nriag_egypt}}
  \and
  % GRANDMA - Shifter + Data analysis + TRT + PWT
  M.~Tanasan\inst{\ref{narit_thailand}}
  \and
  % GRANDMA - TRT
  K.~S.~Tinyanont\inst{\ref{narit_thailand}}
  \and
  % GRANDMA Core Team - KNC
  D.~Turpin\,\orcidlink{0000-0003-1835-1522}\inst{\ref{aim_paris_saclay}}
  \and
  % GRANDMA GTC
  A.~de~Ugarte~Postigo\,\orcidlink{0000-0001-7717-5085}\inst{\ref{aix_marseille_lam}}
  \and
  % EP
  B.T.~Wang\,\orcidlink{0000-0001-9342-1485}\inst{\ref{ynao_china_2},\ref{ucas_beijing}}
  \and
  % TNOT Observer
  L.T.~Wang\inst{\ref{tsinghua_physics}}
  \and
  % TNOT Observer
  X.F.~Wang\,\orcidlink{0000-0002-7334-2357}\inst{\ref{tsinghua_physics}}
  \and
  % GBM
  Z.M.~Wang\,\orcidlink{0009-0006-9824-2498}\inst{\ref{bnu_astro}}
  \and
  % COLIBRI
  A.M.~Watson\,\orcidlink{0000-0002-2008-6927}\inst{\ref{unam_astronomy}}
  \and
  % EP
  H.Z.~Wu\inst{\ref{hust_astronomy}}
  \and
  % EP
  Q.Y.~Wu\inst{\ref{naoc_china},\ref{ucas_astronomy_2}}
  \and
  % EP
  J.J.~Xu\inst{\ref{ihep_china}}
  \and
  % TNOT Observer
  Y.S.~Yan\inst{\ref{tsinghua_physics}}
  \and
  % EP
  H.N.~Yang\inst{\ref{naoc_china}}
  \and
  % EP
  W.~Yuan\inst{\ref{naoc_china},\ref{ucas_astronomy_2}}
  \and
  % EP
  H.S.~Zhao\inst{\ref{ihep_china}}
}

\institute{
    {Aix-Marseille Univ., CNRS/IN2P3, Ctr. de Physique des Particules de Marseille, IPhU (France)\label{aix_marseille_cppm_iphu}}
    \and
    {The Raymond and Beverly Sackler School of Physics and Astronomy, Tel Aviv University, Tel Aviv 69978, Israel\label{tau_physics}}
    \and
    {E.Kharadze Georgian National Astrophysical Observatory, Mt. Kanobili, Abastumani, 0301 Adigeni, Georgia\label{georgian_national_astro_obs}}
    \and
    {Samtskhe-Javakheti State University, Akhaltsikhe, Georgia\label{sjsu_georgia}}
    \and
    {Universit\'e de la Côte d'Azur, Nice, France\label{univ_cote_azur}}
    \and
    {Center for Astrophysics and Space Science (CASS), New York University Abu Dhabi, Saadiyat Island, PO Box 129188, Abu Dhabi, UAE\label{nYu_abudhabi_cass}}
    \and
    {New York University Abu Dhabi, PO Box 129188, Saadiyat Island, Abu Dhabi, UAE\label{nYu_abudhabi}}
    \and
    {Instituto de Astronomía, Universidad Nacional Autónoma de México, Apartado Postal 70-264, 04510 México, CDMX, México\label{unam_astronomy}}
    \and
    {School of Physics and Astronomy, University of Minnesota, Minneapolis, MN 55455, USA\label{umn_physics}}
    \and
    {IJCLab, Univ Paris-Saclay, CNRS/IN2P3, Orsay, France\label{ijclab_orsay}}
    \and
    {IRAP, Université de Toulouse, CNRS, CNES, UPS, France\label{irap_toulouse}}
    \and
    {Aix Marseille Univ., CNRS, CNES, LAM, Marseille, France\label{aix_marseille_lam}}
    \and
    {Department of Applied Physics and Astronomy, and Academy for Astronomy, Space Sciences and Technology, University of Sharjah, Sharjah, United Arab Emirates\label{astro_sharjah}}
    \and
    {Oukaimeden Observatory and High Energy Physics, Astrophysics and Geoscience Laboratory, FSSM, Cadi Ayyad University, Marrakesh, Morocco\label{oukaimeden_obs}}
    \and
    {Dipartimento Interateneo di Fisica, Politecnico di Bari, Via E. Orabona 4, 70125, Bari, Italy\label{poliba_physics}}
    \and
    {Istituto Nazionale de Fisica Nucleare (INFN) — Sezione di Bari, 70126 Bari, Italy\label{infn_bari}}
    \and
    {Mullard Space Science Laboratory - University College London\label{mssl_ucl}}
    \and
    {Department of Physics \& Astronomy, Louisiana State University, Baton Rouge, LA 70803, USA\label{lsu_physics}}
    \and
    {School of Earth and Space Exploration, Arizona State University, PO Box 871404, Tempe AZ 85287, USA\label{asu_earth_space}}
    \and
    {Sorbonne Université, CNRS, UMR 7095, Institut d’Astrophysique de Paris, 98 bis boulevard Arago, 75014 Paris, France\label{iap_paris}}
    \and
    {Institute of Physics and Astronomy, Theoretical Astrophysics, University Potsdam, Haus 28, Karl-Liebknecht-Str. 24/25, 14476\label{potsdam_astro}}
    \and
    {Max Planck Institute for Gravitational Physics (Albert Einstein Institute), Am Mühlenberg 1, Potsdam 14476, Germany\label{aei_potsdam}}
    \and
    {Cahill Center for Astrophysics, California Institute of Technology, MC 249-17, 1216 E California Boulevard, Pasadena, CA, 91125, USA\label{caltech_astro}}
    \and
    {Université Paris Cité, CNRS, Astroparticule et Cosmologie, F-75013 Paris, France\label{apc_paris}}
    \and
    {KNC, Volkssternwarte Paderborn, Im Schlosspark 13, 33104\label{knc_paderborn}}
    \and
    {National Research Institute of Astronomy and Geophysics (NRIAG), 1 El-marsad St., 11421 Helwan, Cairo, Egypt\label{nriag_egypt}}
    \and
    {Xinjiang Astronomical Observatory, Chinese Academy of Sciences, Urumqi, Xinjiang, 830011, China\label{xao_china}}
    \and
    {School of Astronomy and Space Science, University of Chinese Academy of Sciences, Beijing 100049, China\label{ucas_astronomy}}
    \and
    {School of Physics and Astronomy, University of Leicester, Leicester, UK\label{leicester_astro}}
    \and
    {Observatorio de Calar Alto, Sierra de los Filabres, Gérgal, Almería 04550, Spain\label{calar_alto}}
    \and
    {Osservatorio Astronomico di Brera, via E. Bianchi 46, I-23807 Merate (LC), Italy\label{brera_obs}}
    \and
    {KNC, AAVSO, Hidden Valley Observatory(HVO), Colfax, WI. USA\label{colfax_usa}}
    \and
    {Instituto de Radioastronomía y Astrof\'isica, Universidad Nacional Aut\'onoma de M\'exico, Antigua Carretera a P\'atzcuaro \# 8701,\\ Ex-Hda. San Jos\'e de la, Huerta, Morelia, Michoac\'an, M\'exico C.P. 58089, Mexico\label{iraf_unam}}
    \and
    {Instituto de Astronom{\'\i}a, Universidad Nacional Aut\'onoma de M\'exico, km 107 Carretera Tijuana-Ensenada, 22860 Ensenada, Baja California, México\label{unam_ensenada}}
    \and
    {American University of Sharjah, Physics Department, PO Box 26666\label{aus_physics}}
    \and
    {University of Trento, Via Calepina, 14, 38122 Trento TN, Italy\label{univ_trento}}
    \and
    {Purple Mountain Observatory, Chinese Academy of Sciences, Nanjing 210023, China\label{pmo_china}}
    \and
    {School of Astronomy and Space Sciences, University of Science and Technology of China, Hefei 230026, China\label{ustc_astronomy}}
    \and
    {National Astronomical Observatories, Chinese Academy of Sciences, Beijing 100101, China\label{naoc_china}}
    \and
    {School of Astronomy and Space Science, University of Chinese Academy of Sciences, Beijing 100049, China\label{ucas_astronomy_2}}
    \and
    {Institute for Frontier in Astronomy and Astrophysics, Beijing Normal University, Beijing 102206, China\label{bnu_astro}}
    \and
    {FZU - Institute of Physics of the Czech Academy of Sciences, Na Slovance 1999/2, CZ-182 21, Praha, Czech Republic\label{fzu_praha}}
    \and
    {Astrophysics Science Division, NASA Goddard Space Flight Center, Mail Code 661, Greenbelt, MD 20771, USA\label{nasa_goddard}}
    \and
    {Center for Space Sciences and Technology, University of Maryland, Baltimore County, Baltimore, MD 21250, USA\label{umbc_space}}
    \and
    {Center for Research and Exploration in Space Science and Technology, NASA/GSFC, Greenbelt, Maryland 20771, USA\label{crest_nasa}}
    \and
    {KNC, K26 / Contern Observatory (private obs.), 1, beim Schmilberbour, 5316 Contern, Luxembourg\label{knc_contern}}
    \and
    {KNC, Institut Polytechnique des Sciences Avancées: Ivry-sur-Seine, Île-de-France, FR\label{knc_ipsa}}
    \and
    {Physics Department, Tsinghua University, Beijing, 100084, China\label{tsinghua_physics}}
    \and
    {Instituto de Ciencias Nucleares, Universidad Nacional Aut\'onoma de M\'exico, Apartado Postal 70-264, 04510 M\'exico, CDMX, Mexico\label{icn_unam}}
    \and
    {Yunnan Observatories, Chinese Academy of Sciences, Kunming 650216, People’s Republic of China\label{ynao_china}}
    \and
    {Center for Astronomical Mega-Science, Chinese Academy of Sciences, Beijing 100012, People’s Republic of China\label{cams_china}}
    \and
    {Key Laboratory for the Structure and Evolution of Celestial Objects, Chinese Academy of Sciences, Kunming 650216, People’s Republic\label{klo_china}}
    \and
    {Facultad de Ciencias, Universidad Nacional Autónoma de México, Apartado Postal 70-264, 04510 México, CDMX, México\label{fac_ciencias_unam}}
    \and
    {Institute of Astronomy and NAO, Bulgarian Academy of Sciences, 72 Tsarigradsko Chaussee Blvd., 1784 Sofia, Bulgaria\label{inao_bulgaria}}
    \and
    {National Astronomical Research Institute of Thailand (NARIT)\label{narit_thailand}}
    \and
    {KNC, International Astronomical Center, Abu Dhabi, UAE\label{knc_iac}}
    \and
    {W.W. Hansen Experimental Physics Laboratory, Kavli Institute for Particle Astrophysics and Cosmology, Department of Physics and SLAC National Accelerator Laboratory, Stanford University, Stanford, CA 94305, USA\label{stanford_kipac}}
    \and
    {Université de Strasbourg, CNRS, IPHC UMR 7178, F-67000 Strasbourg, France\label{iphc_strasbourg}}
    \and
    {Ulugh Beg Astronomical Institute, Uzbekistan Academy of Sciences, Astronomy str. 33, Tashkent 100052, Uzbekistan\label{ulugh_beg_inst}}
    \and
    {KNC, AAVSO Observer; Pobedy Street, House 7, Flat 60, Yuzhno-Morskoy, Nakhodka, Primorsky Krai 692954, Russia\label{knc_aavso}}
    \and
    {Main Astronomical Observatory of the National Academy of Sciences of Ukraine, 27 Akademik Zabolotny St., Kyiv, 03143, Ukraine\label{mao_ukraine}}
    \and
    {KNC, Société Astronomique de France, Observatoire de Dauban, FR 04150 Banon, France\label{knc_saf}}
    \and
    {Astronomical Observatory, Taras Shevchenko National University of Kyiv, 3 Observatorna St., 04053 Kyiv, Ukraine\label{kyiv_astro_obs}}
    \and
    {Department of Physics, University of Aberdeen, Aberdeen AB24 3UE, UK\label{aberdeen_physics}}
    \and
    {Université Paris-Saclay, Université Paris Cité, CEA, CNRS, AIM, 91191, Gif-sur-Yvette, France\label{aim_paris_saclay}}
    \and
    {Yunnan Observatories, Chinese Academy of Sciences, 650216 Kunming, Yunnan Province, People’s Republic of China\label{ynao_china_2}}
    \and
    {University of Chinese Academy of Sciences, Beijing 100049, China\label{ucas_beijing}}
    \and
    {Department of Astronomy, Tsinghua University, Beĳing 100084, China\label{tsinghua_astronomy}}
    \and
    {Department of Astronomy, School of Physics, Huazhong University of Science and Technology, Luoyu Road 1037, Wuhan 430074, China\label{hust_astronomy}}
    \and
    {Key Laboratory of Particle Astrophysics, Institute of High Energy Physics, Chinese Academy of Sciences, Beijing 100049, China\label{ihep_china}}
}

\date{Received xx xx, xx; accepted xx xx, xx}

% \abstract{}{}{}{}{} 
% 5 {} token are mandatory
 
  \abstract
   {Gamma-Ray Burst GRB~241030A ($z = 1.411$) exhibited a particularly bright afterglow (similar to the ``BOAT'', GRB~221009A), detected across gamma-ray, X-ray, UV, and optical bands. The extensive, multi-wavelength observations of this remarkable event provide a valuable opportunity to advance our understanding of GRB afterglow physics.}
   {We aim to constrain the physical properties of the jet, its microphysics, and the characteristics of the circumburst environment in the context of forward-shock emission}%; as well as characterize the potential host galaxy of the GRB.}
   {We compiled multi-wavelength observations spanning from a minute to a week after the prompt emission, processing the data through a unified photometry pipeline. Leveraging this comprehensive dataset, we analysed the observations both analytically and using Bayesian inference with two independent models. Our models assume that the afterglow emission arises from the strong forward shock of a laterally structured jet, with possible contributions from synchrotron self-Compton (SSC) scatterings.}
   %The analysis of the afterglow observations was conducted using an analytical approach and two independent Bayesian models that predict afterglow emission behind a strong forward shock.} 
   %corrected for the line-of-sight extinction and used a coherent photometry pipeline on the eleven optical instruments involved, together with the three gamma-ray instruments at play. The analysis of the afterglow observations was conducted using an analytical approach and two independent Bayesian models that predict afterglow emission behind a strong forward shock. }
   {We find that our models do reproduce the afterglow observations accurately, from the X-rays to the optical, favouring a jet propagating into a constant-density interstellar medium, with a viewing angle within the jet core. However, both analyses -- with and without the inclusion of SSC scatterings -- require parameter values that are extreme compared to expectations from standard theory. In particular, our results imply extremely energetic jets despite regular prompt energy, leading to a very inefficient prompt emission. Furthermore, the jets are particularly inefficient at accelerating particles, with low $\epsilon_\mathrm{e}$ and $\epsilon_\mathrm{B}$, leading to significant SSC emission. Finally, our analyses indicate that the jets have large opening angles and propagate in high-density media.}
   {If the afterglow is indeed powered by radiation emitted behind a strong forward shock, our results place GRB~241030A within a sub-class of GRBs characterised by extreme kinetic energies, large jet opening angles, and very low prompt emission efficiencies, below $10^{-3}$, with strong SSC radiation. These predictions are difficult to reconcile with typical expectations from other GRBs. We therefore suggest that the afterglow of GRB~241030A is not solely powered by forward shock emission, and we discuss other options such as a long-lasting reverse-shock contribution.}

   \keywords{(Stars:) Gamma-ray burst: general - (Stars:) Gamma-ray burst: individual: GRB241030A - Radiation mechanisms: non-thermal – Methods: statistical - Techniques: photometric }

   \maketitle
%
%-------------------------------------------------------------------

\section{Introduction}

Gamma-ray bursts (GRBs) are among the most energetic and luminous events occurring in the Universe. They are typically divided into two broad categories based on the duration of their gamma-ray emission and the hardness of their spectra: long-duration bursts, which last more than two seconds and have ‘soft’ spectra, and short-duration bursts, which last less than two seconds and have ‘hard’ spectra. Long GRBs are generally associated with the core collapse of massive stars, events that are often referred to as ‘collapsars’, while short GRBs are due to the merger of a neutron star with another neutron star or a black hole (e.g. \citealt{2017AdAst2017E...5C,2017ApJ...848L..13A}). At least one (GRB~170817A) has been associated with gravitational waves \citep{2017ApJ...848L..13A, 2017ApJ...848L..14G}. In both long and short GRBs, highly relativistic jets are launched from the vicinity of the newly formed compact object. As these jets propagate outward and interact with the surrounding environment, they generate shocks which allow for particle acceleration and give rise to the so-called afterglow emission, from radio to very high energies \citep{1998ApJ...497L..17S,1999ApJ...520..641S,1999ApJ...519L..17S,2004RvMP...76.1143P}.

GRB~241030A is a long GRB detected at $t_0$ = 05:48:03 UT (used in the following as reference $t_0$) on 2024 October 30 ($T_{90}=166$\,s in 50--300 keV) by the \textit{Fermi} Gamma-ray Burst Monitor (GBM, \citealt{2024GCN.38015....1D}). It was also detected by many other gamma-ray detectors. An observational campaign including several follow-up facilities led to the discovery of the X-ray and optical counterparts and is summarized in Appendix~\ref{Appendix:campaign}. \red{Spectroscopic observations provided a redshift measurement of $\mathrm{z}=1.411$ \citep{Zheng2024_GCN37959, 2024GCN.38027....1L}.}

%This broadband afterglow emission is explained as the relativistic ejecta interacts with the external medium, shifting characteristic frequencies from higher to lower energies over time. 
Previous studies have shown a moderate correlation between the afterglow brightness and the isotropic equivalent energy of the prompt gamma-ray emission $E_{\rm iso}$ \citep{2008ApJ...689.1161G, 2009ApJ...701..824N, 2010ApJ...720.1513K, 2011ApJ...734...96K}. The case of GRB~241030A offers an opportunity to explore these connections further, to probe the structure and environment of the GRB jet and test the limits of models where only synchrotron emission is accounted for.

In the standard model for GRB afterglows, the GRB jet interacts with the surrounding medium, forming a relativistic forward shock behind which the magnetic field is amplified and electrons are accelerated. This general model, first proposed in the late 1990s, successfully explains the broad-band, smoothly decaying light curves observed in many GRBs \citep{1997ApJ...476..232M, 1998ApJ...497L..17S, 1999PhR...314..575P}. These accelerated electrons emit synchrotron radiation, whose spectral and temporal evolution depends on microphysical parameters (e.g. $\epsilon_e$, $\epsilon_B$, the fractions in the electrons and magnetic field of the total internal energy in the shocked material behind the shock, and $p$, the power-law index of the energy distribution of the electrons) and the density profile of the circumburst medium $n(r)$, where $r$ is the radial distance to the source, of the surrounding medium.

Two main types of circumburst media are typically considered: a uniform interstellar medium (ISM) and a stellar wind environment with a density decreasing as $r^{-2}$. The ISM model is appropriate for GRBs occurring in constant-density regions, potentially short GRBs associated to mergers which have migrated in their host galaxy away from star-forming regions, while the wind model is consistent with progenitors such as Wolf-Rayet stars that eject strong stellar winds prior to collapse \citep{1998ApJ...499..301C, 2000ApJ...536..195C}. The evolution of the afterglow flux and spectral breaks (for example, the cooling frequency, $\nu_\mathrm{c}$) differs significantly between these environments, and distinguishing between them provides insight into the nature of the progenitor and its surroundings.

%More sophisticated models incorporate the effects of jet collimation. As the jet slows and the relativistic beaming angle widens, observers begin to detect the edge of the jet. This produces a steepening in the light curve known as a ‘jet break’ \citep{1997ApJ...487L...1R,1999ApJ...525..737R, 1999ApJ...519L..17S}. The timing and shape of this break yield estimates of the jet opening angle and hence the beaming-corrected energy, which is crucial for understanding the true energetics of GRBs. \cp{\textit{I'm not too sure about this first part: jet collimation is a standard ingredient of afterglow modelling, and I don't know if it is worth putting that much emphasis on the jet break in the introduction. I would either remove this first part or make it much shorter and maybe focus on the prompt efficiency (ratio of prompt to afterglow energy) which is an uncertain quantity in GRB physics.}} 
Furthermore, reverse shocks propagating into the ejecta can produce early-time optical flashes, providing further diagnostics of the outflow properties \citep{1999ApJ...520..641S, 2000ApJ...545..807K}. Synchrotron photons can be upscattered to higher energies by the seed electron population, a process known as synchrotron self-Compton (SSC) radiation. This can power the very high energy (VHE) emission detected in some GRB afterglows, such as GRB~190114C \citep{2019Natur.575..455M, 2019Natur.575..459M} or GRB~221009A \citep{LHAASO_221009A}. Under certain conditions (e.g. a very low magnetization), SSC can significantly alter the synchrotron spectrum across all wavelengths and should therefore be included in afterglow models, even in the absence of VHE detection \red{\citep{2001ApJ...548..787S, 2009ApJ...703..675N, 2024ApJ...970..135M, Pellouin:2024gqj, 2025MNRAS.538..281H}}.

Further research advances extended afterglow modeling to include structured jets, energy injection, and magnetization. In structured jet models, the energy per solid angle varies with the angle from the jet axis \red{(see early works by e.g. \citealt{1998ApJ...499..301M, 2003ApJ...592..390P})}, which can account for the observed diversity in afterglow behavior and explain events observed off-axis, such as GRB 170817A \red{\citep{2017ApJ...850...24L,2017Natur.551...71T,2017Sci...358.1579H, 2018NatCo...9..447L,2018MNRAS.478L..18G,2018MNRAS.479.1578G}}. %Energy injection, possibly due to prolonged central engine activity or stratified ejecta, can flatten light curves \cp{\textit{What does that mean? Are we referencing X-ray plateaus?}} or produce rebrightenings \citep{2005ApJ...627..877N, 2006ApJ...642..354Z}. Magnetized outflows introduce further complexity, altering the efficiency of reverse shocks and affecting polarization signatures \citep{2003ApJ...597..455Z, 2005ApJ...628..315F}.

In the case of GRB~241030A, interest in following this burst quickly emerged because its afterglow appeared very bright in UV/optical in the initial observations. Its well-sampled afterglow across the X-ray, optical and infrared bands provides an excellent opportunity to test afterglow models. The temporal and spectral evolution of the emission may help discriminate between an ISM or wind-like environment, constrain the jet geometry, and evaluate the role of energy injection or jet structure. %Any deviations from the standard synchrotron model, such as plateaus, rebrightenings, or chromatic breaks, may offer insight into the nature of the central engine or the presence of angular or radial jet stratification. As such, GRB~241030A serves as a valuable case study for examining the diversity and limitations of current afterglow theories. \cp{\textit{I'm confused, in this paragraph we focus on 241030A and mention general things about afterglows which are not all relevant for this case (there is no plateau, no chromatic breaks, no rebrightening after 400s, and we do not test energy injetion or radial stratification). We should either let this paragraph be general, or remove aspects that are not relevant to 241030A.}}
Yet, \citet{2025ApJ...987..129W} have investigated the prompt thermal X-ray emission component and origins of the very early UVOT clear and $U$ band emission. Using early \textit{Swift}-XRT and \textit{Swift}-UVOT observations of GRB~241030A, \cite{2025ApJ...987..129W} found strong optical variability that traces the gamma-ray activity, possibly from internal shocks. A thermal component, likely photospheric, was found with a temperature of a few keV and an inferred Lorentz factor between 20 and 80. A sharp rise in the $U$-band suggested the onset of external shock emission. Thus, we aim to complement these investigations by looking at the multi-wavelength afterglow observation from a minute to 9.5~days after the prompt emission. Additional studies of the prompt emission have been presented in \citet{2025arXiv251017323W, 2025arXiv251024864V}, which analyse the gamma-ray and early X-ray emission of GRB~241030A.
%Thus, we aim to complement these investigations with a revised version of UV data, and additional observations in hard and soft X-rays and optical especially at longer term. \cp{\textit{Do we actually do it? Maybe we should simply say that we complement these investigations by looking at the multi-wavelength afterglow}}

In Section~\ref{section:observations}, we present the multi-wavelength (gamma-ray to optical) afterglow observations of GRB~241030A. In Section~\ref{section:host_galaxy}, we determine the line-of-sight extinction and leverage our optical follow-up observations to search for the host galaxy. Section~\ref{section:MWafterglowanalysis} is dedicated to the multi-wavelength study of the afterglow of GRB~241030A, using an analytical approach and Bayesian inference with two different models. Finally, in Section~\ref{sec:discussion}, we discuss our findings and conclude this work in Section~\ref{sec:conclusion}.

\section{Multi-wavelength observations of GRB~241030A}\label{section:observations}

%In this section we present the multiwavelength dataset of GRB241030A afterglow, as well as the dataset of the host galaxy. (Redundant: already in the title of the section.)
The gamma-ray \& X-ray observations used in the following analysis are presented in Appendix~\ref{ap:data}, Table~\ref{tab:grandmadata_he} and shown in Fig.~\ref{fig:grandma+community}. The UV/optical observations used in the analysis are presented in Appendix~\ref{ap:data}, Tables~\ref{tab:op-1},\ref{tab:op-2},\ref{tab:op-3},\ref{tab:op-4},\ref{tab:op-5} and are also shown in Fig. \ref{fig:grandma+community}. The optical data points are also accessible in the \href{https://skyportal-icare.ijclab.in2p3.fr/public/sources/GRB241030/version/83fbc543fcc4de72247b762986a87500}{Skyportal page of the event}. %The host galaxy dataset is presented in Table \ref{table:hostdata}.

%\subsection{GBR 241030A detection} \label{subsec:30A-gen-info}

%First alert and XRT Afterglow detection, localization, flux decline, redshift

%At 05:48:03 UT, the Swift Burst Alert Telescope (BAT) triggered on GRB~241030A and immediately located the burst at RA(J2000) = 22h 52m 08s, Dec(J2000) = +80d 26' 20" with a 3-arcmin uncertainty. This GRB was also detected by FERMI GBM at 05:48:03 UT on 30 Oct 2024. The BAT light curve showed a complex structure with a duration of at least 180 seconds and a peak count rate of ~12000 counts/sec at ~162 seconds post-trigger. The XRT began observations at 05:49:16.6 UT, finding a bright X-ray source at RA(J2000) = 22h 52m 33.60s, Dec(J2000) = +80d 26' 53.5" with a 4.7-arcsecond uncertainty, located 72 arcseconds from the BAT position. The UVOT took a 150-second exposure 82 seconds after the BAT trigger, detecting a candidate afterglow at RA(J2000) = 22h 52m 33.57s, Dec(J2000) = +80d 26' 59.9" with a 0.61-arcsecond uncertainty. The estimated magnitude is 15.42 (with a 1-sigma error of 0.14). No extinction correction was applied. 

\subsection{Prompt: E$_\mathrm{iso}$ computation}
\label{section:E_iso_computation}

We calculated the total isotropic-equivalent energy released in gamma-rays using measurements from both \textit{Fermi}-GBM and \textit{Konus}-Wind. \citet{2024GCN.37982....1R} report the \textit{Konus}-Wind measurement of $E_\mathrm{iso} = (2.84^{+1.04}_{-0.68}) \times 10^{53}$~erg. 
For \textit{Fermi}-GBM we utilize the initially reported time-integrated spectral results from the \textit{Fermi}-GBM team \citep{2024GCN.38015....1D}. 
This corresponds to a fluence of $(6.2 \pm 0.1) \times 10^{-5}$~erg $\cdot$ cm$^{-2}$, from a best-fit Band function with a peak energy 
$E_\mathrm{peak} = (130 \pm 10)$~keV, low-energy index $\alpha = -1.35 \pm 0.02$, and high-energy index $\beta = -2.3 \pm 0.1$. 
Accounting for the cosmological $k$-correction \citep{Bloom2001A}, we calculate 
$E_\mathrm{iso} = (4.0 \pm 0.1) \times 10^{53}$~erg. 
We also extract time-resolved spectra from the Fermi-GBM data, and by combining the fluxes from the best-fit Band function in each interval, we yield a consistent
$E_\mathrm{iso} = (4.4 \pm 0.2) \times 10^{53}$~erg. 
A similar order of magnitude $E_\mathrm{iso} = (3.3 \pm 0.1) \times 10^{53}$~erg was also reported by \cite{2025arXiv251024864V}, employing smoothly broken powerlaw models with high-energy cutoff. 

\subsection{Early afterglow with \textit{Swift}}

After slewing to the burst, \textit{Swift}-UVOT started collecting data at approximately $t_0 + 80~\mathrm{s}$, while the GRB was still in its prompt phase ($T_{90} \sim 200~\mathrm{s}$). In the white band, the emission appears to rise until around $t_0 + 120~\mathrm{s}$, at which point it peaks (with an AB mag of about 15.6), then fades shortly after.  At about $t_0 + 200~\mathrm{s}$, the emission starts to rise again.  In the $u$~band, the emission rises again until about $t_0 + 400~\mathrm{s}$, peaking at 13.6 mag (AB), after which point it begins to fade again following a power-law decay. Thus, we refer to times before $t_0 + 400~\mathrm{s}$ as the prompt emission, and times after $t_0 + 400~\mathrm{s}$ as the afterglow.

In X-rays, during the prompt emission phase a very bright flare can be seen from about 80~s to 430~s post-trigger. The flare exhibits substantial variability, but the average brightness peaks at around 180~s.  After 430~s, the X-ray emission begins to follow a power-law decay. %We analyse the X-ray afterglow data only and our results are reported in Appendix~\ref{sec:afterglowXray}. 
A joint analysis of the X-ray and gamma-ray (\textit{Swift}-BAT) data by \citet{2025ApJ...987..129W} found the presence of a thermal component with a temperature of a few keV during the flare period, which the authors interpreted as the photosphere radiation. They also attribute the $u$-band peak to the onset of the external shock emission.

\subsection{Afterglow: Gamma-ray data}

The burst was detected by the \textit{Fermi}-LAT~\citep{2009ApJ...697.1071A} in the 100~MeV--100~GeV range \citep{LAT_GCN}, occurring $19^{\circ}$ from the LAT boresight at the time of the GBM trigger ($t_0$) and remaining in the Field of View (FoV) for $\sim 800 \mathrm{s}$. In this time interval we performed a standard LAT data analysis, following the procedure described in \citet{2cat}, and implemented with the ``Multi-Mission Maximum Likelihood`` (\texttt{ThreeML}, \citealt{3ML}). We detected 9 high-energy photons with more than 90\% probability of association with the GRB. The highest-energy photon detected was a 2 GeV event at $t_0$ + 470~s. Therefore, the analysis was restricted to the 100 MeV–10 GeV range. The \textit{Fermi}-LAT events were binned, with each bin containing at least three significant photons and a detection with at least $3 \sigma$ significance. For each time interval, using the \texttt{FermiLATLike} plugin, we conducted an unbinned maximum likelihood fit of the energy spectrum using a simple power-law (PL) model, $F(E) = F_0 \times \Big( \frac{E}{E_0} \Big)^{-\gamma}$, where the flux normalization $F_0$ and the photon index $\gamma$ were treated as free parameters. The resulting light curve is shown in Fig.~\ref{fig:grandma+community} (black dots) and the data is provided in Table~\ref{tab:grandmadata_he}. The GRB position re-enteres the LAT FoV from about 5 ks and until about 13 ks. Standard analysis resulted in no detection and we computed upper limits. We searched for an extended emission covering the whole afterglow period observed by the other instruments (until 700 ks), but no significant high-energy emission was detected. More details are available in Appendix~\ref{sec:lat_appendix}.

\subsection{Afterglow: X-ray data}

\begin{figure*}
    \centering
    \includegraphics[width=1\textwidth]{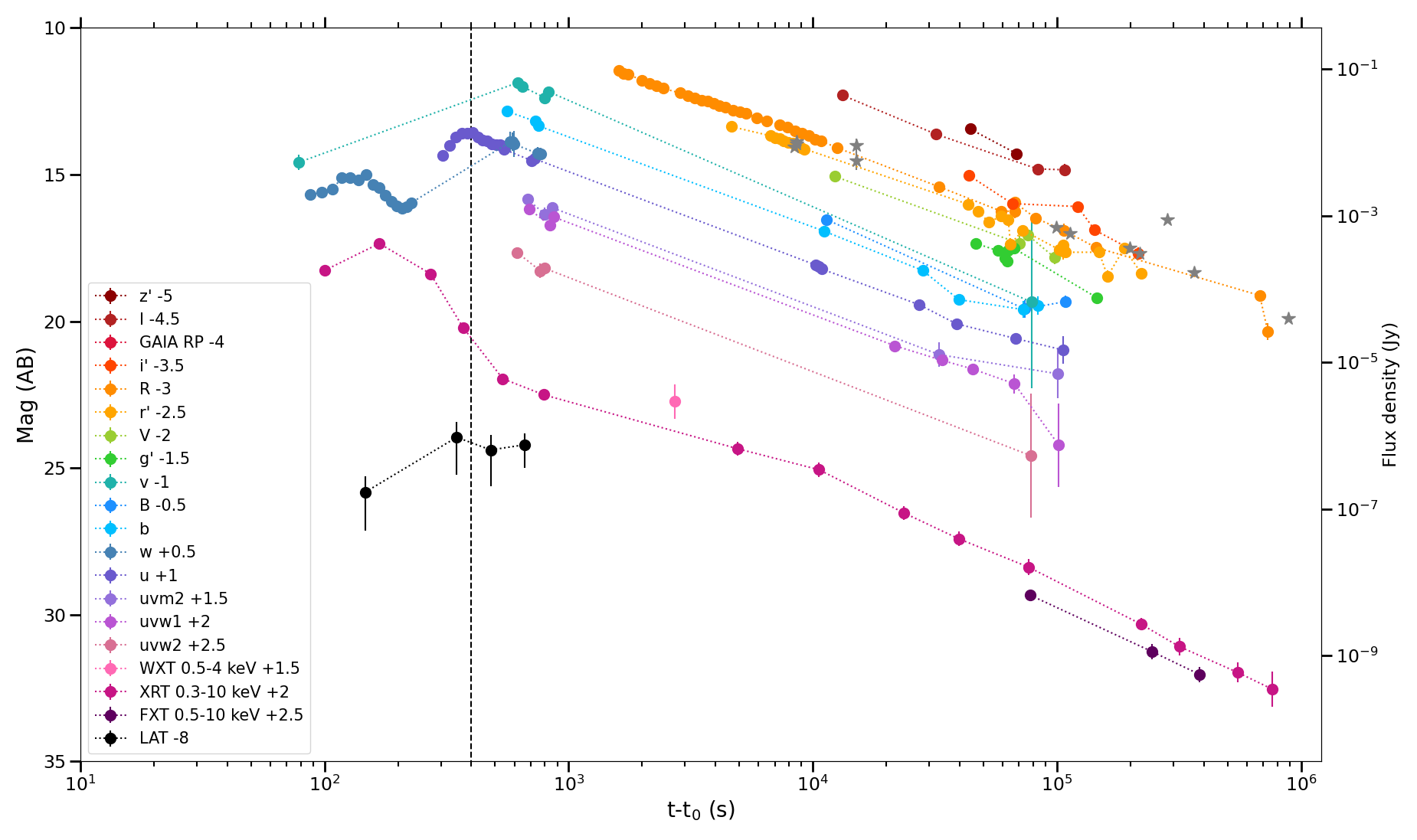}
    \caption{Light curves of GRB~241030A, corrected for Galactic and host extinction. Observations after the vertical dashed line (shown at 400 s post T0) were used for the afterglow analysis in Section~\ref{section:MWafterglowanalysis}. Grey stars are GRB~221009A {r'} band magnitude from \citet{2023ApJ...948L..12K}, adjusted to redshift 1.411 with k-correction, -2.5 magnitude for comparison with the GRB~241030A {r'} light curve (see Section~\ref{sec:bayesian_syn}). }
    \label{fig:grandma+community}
\end{figure*}

We analysed X-ray data from \textit{Swift}-XRT, \textit{EP}-WXT, and \textit{EP}-FXT. First, the XRT light curve (0.3--10~keV) of GRB~241030A was acquired from the UK Swift Science Data Center and extracted from the burst analyser\footnote{\url{https://www.swift.ac.uk/burst_analyser/}} \citep{2007A&A...469..379E, 2009MNRAS.397.1177E}. We performed a re-binning of the XRT light curve by dividing the observations into fifteen time windows in the same way as in \citet{2023ApJ...948L..12K}.
For each time window we computed the mean value and standard deviation to produce single data points. The light curve is presented in Fig. \ref{fig:grandma+community} and the data is provided in Table~\ref{tab:grandmadata_he}.

Secondly, we analysed observations from \textit{EP}-WXT \citep{Cheng_2025} from 2715 seconds after $t_0$ with an exposure of approximately 1.2 ks. The light curve exhibited no significant variability throughout the observation. The spectrum was fitted in \texttt{XSPEC} by an absorbed power law model \textit{tbabs*ztbabs
*powerlaw} \citep{1999ascl.soft10005A}, where a fixed redshift of 1.411 was used, and the column density was fixed at the Galactic value of $N_\mathrm{H}^\mathrm{Gal}=$ 1.79 $\times \ 10^{21}$ cm$^{-2}$. It should be noted that the flux derived from WXT corresponds to the 0.5--4~keV energy band. Subsequently, we performed three follow-up observations with the FXT \citep{Chen2021SPIE} in the 0.5--10~keV band. For the spectral fitting, we applied the same model used for the WXT data. Due to the low photon statistics in the last FXT observation, we combined data from the final two observational epochs to perform a joint fit. We provide the results in Table~\ref{tab:grandmadata_he} and include the light curve in Fig. \ref{fig:grandma+community}. %The fitted results and corresponding fitting statistics are by shown in Table \ref{tab:ep_spec}, the light curve is presented in Fig. \ref{fig:grandma+community}.

%\begin{table*}
%\centering
%\scriptsize
%\begin{threeparttable}
%\caption{\textbf{The fitting results and corresponding fitting statistics for the afterglow spectra.} All errors represent the 1$\sigma$ uncertainties.}
%\label{tab:ep_spec}
%\small
%\begin{tabular}{cccccccc}
%\toprule
%Instruments & Observation Start Time(UTC) & Time Interval  & Flux \tnote{1} & $%\Gamma$ & CSTAT/(d.o.f) \\
%& & (second) & ($\rm erg\,cm^{-2}\,s^{-1}$) &   & \\
%\hline
%WXT& 2024-10-30T06:33:18 & 1166 & $1.1 ^{+0.3}_{-0.2} \times 10^{-10}$  & 2.6$^{+0.9}_{-0.3}$ &27/30\\

%\midrule
%FXT& 2024-10-31T03:26:35 & 3786 & $1.5 ^{+0.1}_{-0.1} \times 10^{-12}$   & 2.2$^{+0.1}_{-0.1} & 213/205 \\
% & 2024-11-02T01:58:27 & 3173  &  $2.7 ^{+0.6}_{-0.4}\times 10^{-13}$ & 2.1$^{+0.4}_{-0.4}$ & 98/83\\
%  & 2024-11-03T16:48:20 & 6593 &  $1.3 ^{+0.3}_{-0.2} \times 10^{-13}$  &  & \\
%\bottomrule
%\end{tabular}
%\begin{tablenotes}
%\footnotesize
%
%\end{tablenotes}
%\end{threeparttable}
%\end{table*}

\subsection{Afterglow: UV/Optical data}
 \label{afterglowobs}
The \textit{Swift}-UVOT data was analysed using the standard UVOT tools included with HEASoft (v6.34).  
Most of the data from the first orbit (i.e., data collected before 2024-10-30 06:03 UTC) were taken with UVOT operating in Event Mode (in which each event is time-tagged with a resolution of 11 ms), so we used {\tt uvotevtlc} to create background-subtracted light curves from the event data, using a time binning of 20 seconds.  
For the rest of the data, collected in Imaging Mode, we calculated the magnitudes using the standard tool {\tt uvotsource} for all filters (white, $v$, $b$, $u$, $uvw1$, $uvm2$, and $uvw2$).
Due to the source's high brightness during the first orbit, we used an $r=15''$ source extraction radius (i.e., aperture), and an $r=6''$ extraction radius for all subsequent data.
We used two source-free regions ($r=20''$ and $r=45''$) within 1.5$'$ of the burst position for the background regions (required input by {\tt uvotsource} and {\tt uvotevtlc} for background subtraction).

Besides, we processed in a coherent approach all the optical images from GRANDMA, KNC and COLIBRÍ together with images from TRT-SRO $R$ band presented by \citet{2025ApJ...987..129W}.
%The optical images presented in this study from GRANDMA, KNC and Colibri together with images from TRT-SRO $R$ band presented by \citet{2025ApJ...987..129W} were all processed in a coherent approach. \cp{\textit{Rephrase the first sentence.}}
All images in clear, $GAIA-RP$, $R$, $V$, $I$, $B$, $g'$, $r'$, $i'$, and $z'$ underwent telescope-specific pre-processing, including bias, dark subtraction, and flat-fielding. For some images, co-addition was performed using the \texttt{SWarp} software \citep{2010ascl.soft10068B} to enhance signal-to-noise ratio, and astrometric solutions were acquired using Astrometry.net \citep{2010astrometrynet}. All data have been reduced by a single data processing pipeline, STDPIPE\footnote{\raggedright{\url{https://gitlab.in2p3.fr/icare/stdpipe}} } \citep{stdpipe,Karpov_2025}, a python package that uses pre-configured scripts for image analysis. Specifically, we used the online implementation of the package called STDweb\footnote{available at \url{http://stdweb.favor2.info/}}. For photometric calibration, the catalogs were chosen based on the filters used in the images. The images acquired with sloan filters were calibrated using the Pan-STARRS DR1 catalog \citep{2016arXiv161205560C}. Images observed with Johnson-Cousins filters were calibrated using the Gaia DR3 Syntphot catalog \citep{2023A&A...674A..33G}, which offers synthetic photometry from low-resolution (XP) spectra of stars observed by Gaia. Lastly, images observed without a filter cover a broad wavelength range, making standard calibration difficult. Hence, we used the $G$-band from the Gaia eDR3 catalog \citep{2023A&A...674A..33G} to handle these images.
For the analysis below, we corrected UV/Optical observation for Galactic extinction using the approach of \citet{Schlafly11}.

\section{Host galaxy extinction and characterisation}
\label{section:host_galaxy}

\subsection{Line-of-sight extinction}\label{sec:los_extinction}

To estimate the line-of-sight host galaxy dust extinction, we created an X-ray-to-optical spectral energy distribution (SED) with \textit{Swift}-XRT and optical data in the \textit{u,b} bands from \textit{Swift}-UVOT; \textit{B, R, I} from TRT and \textit{r} from COLIBRÍ at a common time of $t - t_0 \sim 3~\mathrm{h}$ after the burst trigger. The optical data was interpolated to this mid-time and the time-sliced X-ray spectrum was acquired from the automated data products provided by the public \textit{Swift}-XRT repository \citep{2007A&A...469..379E}.

\begin{figure}
  \centering
  \includegraphics[width=0.95\linewidth]{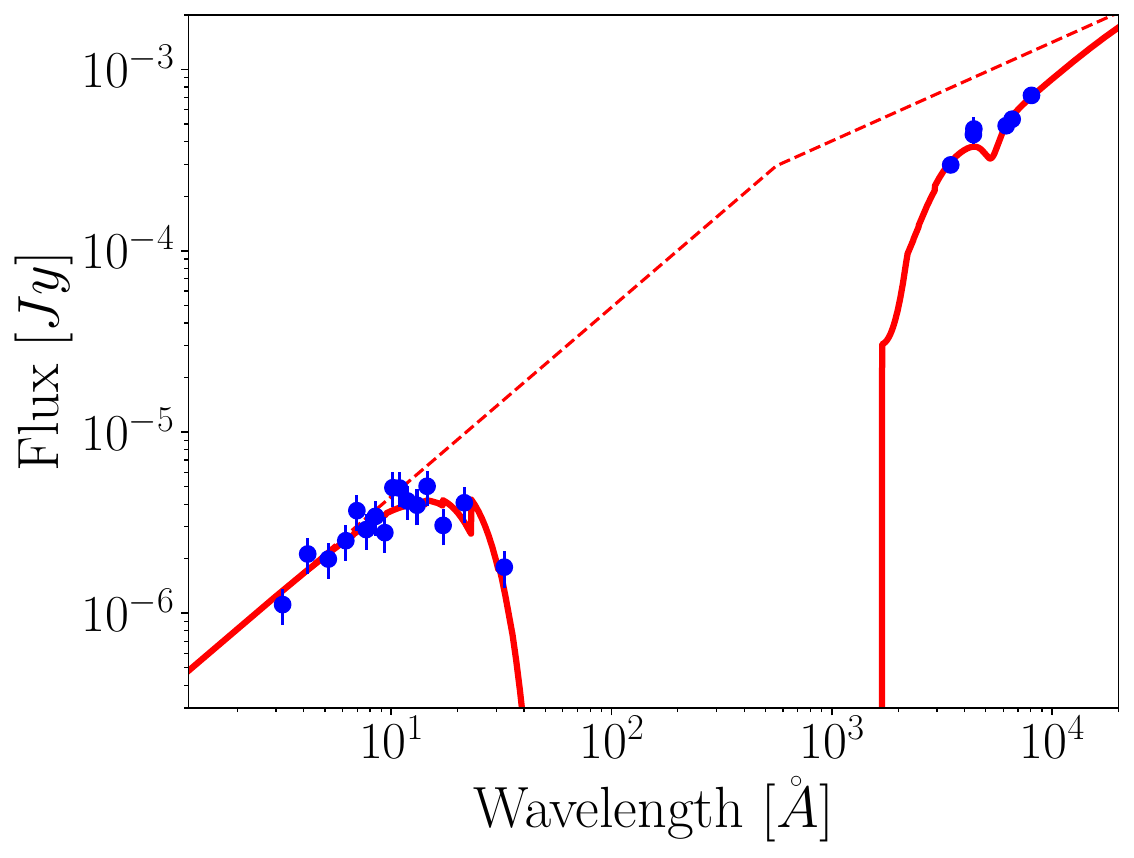}
  \caption{SED of the afterglow of GRB~241030A at $t - t_0 \sim 3~\mathrm{h}$. The dashed line corresponds to the intrinsic broken power law model. The solid line represents the best fit to the data, and includes the absorption in the X-ray band.}
  \label{fig:sedfitextinction}
\end{figure}

We performed the SED fit by following a standard routine using the typical extinction curves of the Milky Way (MW) and the Large (LMC) and Small Magellanic Clouds (SMC) prescribed in \cite{pei_interstellar_1992}. The intrinsic X-ray-to-optical spectrum was modeled with a single or broken power law according to the afterglow theory \citep{1998ApJ...497L..17S}. %The slope of the X-ray wavelengths was assumed to be 0.5 steeper than the spectral slope below the cooling break of the intrinsic spectra, i.e. $\Delta\beta = \beta_\mathrm{X} - \beta_\mathrm{o} = 0.5$. \cp{\textit{This last sentence is not clear, should be rephrased.}} 
According to the standard afterglow model, \red{in the slow cooling regime} the spectral index above the cooling frequency is expected to be steeper by 0.5 than that below it \red{(see Section~\ref{section:empirical_fit})}. We therefore fixed the X-ray slope to be 0.5 larger than the optical one, i.e. $\Delta\beta = \beta_\mathrm{X} - \beta_\mathrm{o} = 0.5$. By fixing the redshift and the Galactic foreground absorption of $N_{\mathrm{H}}^{{\mathrm{Gal}}} = 1.79~\times 10^{21} \; \mathrm{cm}^{-2}$ \citep{collaboration_hi4pi_2016,willingale_calibration_2013}, the best fit is given by a broken power law with a MW extinction curve (Fig. \ref{fig:sedfitextinction}). The best fit parameters are $\beta_\mathrm{X} = \beta_\mathrm{o} + 0.5 = 1.045 \pm 0.062$ (E$_\mathrm{break}$ = 0.022 $\pm$ 0.016 keV), and $E(B-V)$ $<$ 0.195 mag at 3-$\sigma$ confidence level with $\chi^2/\text{d.o.f}$ = 19.057/17. The fit with the LMC extinction curve also gives a reasonable fit ($\chi^2/\text{d.o.f}$ = 20,077/17) yielding $\beta_\mathrm{X} = \beta_\mathrm{o} + 0.5 = 1.045 \pm 0.059$ (E$_\mathrm{break}$ = 0.043 $\pm$ 0.029 keV), and $E(B-V)$ $<$ 0.102 mag. This is compatible with standard values observed for long GRBs at this redshift (e.g. \citealt{2012A&A...537A..15S, 2024A&A...690A.373R}). However, with the lack of NIR data, this amount of extinction might be underestimated. The process of determining optical extinction is more complex without NIR data points, which are less affected by dust extinction than the UV and blue bands. In addition, they can play a role as an anchor for the determination of the intrinsic spectral shape of the afterglow. As a further test, we performed the SED fit at two different epochs, i.e. $t - t_0 \sim 0.19~\mathrm{h}$, using $uvw2,uvm2,uvw1,u,b,v$-bands and $\sim22$~h, with $uvw2,uvm2,uvw1,u,b,v,g,r,i$, in the same manner as previously. The resulting values of the extinction remained consistent with this limit of $E(B-V)$ $<$ 0.195 mag. In light of this analysis, we used in the following an $E(B-V)$ $=$ 0.1 mag to correct our observations from the host extinction.

\subsection{Host galaxy study}\label{sec:host}

We first used the long-term follow-up observation from COLIBRÍ to stack all the images acquired from June 6 to June 10, 2025 (from 219 to 223 days post-trigger) with a total of about 21.43 hours of images. This allowed us to get a deep image of the field at a time when the afterglow contribution was low and to search for a potential host galaxy. In this image, we identified an extended object near the afterglow position at RA = 343.1419, Dec = 80.4496 (deg), which we initially considered as a putative host galaxy of the GRB. Thus, we requested observation with CFHT-MegaCam in $r$ and $i$ bands and with GTC-HiPERCAM in $u$, $g$, $r$, $i$, and $z$ bands. The putative host was well detected in these observations, but these precise measurements of the galaxy allowed us to measure an offset of $1.74 \pm 0.18$ arcsec with the afterglow position. Assuming the \citet{Planck18}'s cosmological parameters, at \red{GRB~241030A's} redshift $z=1.411$, this would represent a physical offset of $15.10 \pm 1.56$ kpc. This value being significantly larger than typical long GRB offsets (e.g. \citealt{Blanchard2016}), we discarded this galaxy as candidate host. To confirm this rejection, we performed the SED fitting of this galaxy using CIGALE\footnote{http://cigale.lam.fr} (Code Investigating GALaxy Emission; \citealt{Boquien2019}), adopting a parameter space similar to \citet{Corre2018}, keeping the redshift as free parameter in the range [0.4,1.5]. We hence obtained a photometric redshift of $z = 0.50^{+0.19}_{-0.10}$. This redshift is compatible with absorption lines detected at lower redshift in the afterglow spectrum \citep{Zheng2024_GCN37959}. We hence conclude it is not the host of GRB~241030A but a foreground galaxy. No other sources are detected in CFHT and GTC images near the afterglow position down to magnitudes $i<25.62$ and $r<25.82$. The non-detection of a host at $z = 1.411$ is consistent with previous long GRBs at similar redshifts, for which hosts were found to be faint ($r > 25$–26 mag) or undetected even in deep observations (e.g. \citealt{Cenko2008,Greiner2015,Perley2016,Perley2016b}). Our limiting magnitudes thus do not exclude the presence of a typical faint GRB host, especially if moderately dust-obscured.

\section{Multi-band afterglow modeling}\label{section:MWafterglowanalysis}

\subsection{Empirical Light curve analysis}
\label{section:empirical_fit}

%\begin{figure}
%    \centering
%    \includegraphics[width=\linewidth]{figures/empirical_LC_not_final.png}
%    \caption{Empirical light curve of GRB241030A from Infarad to X-ray range. A clear estimation of the host galaxy extinction has performed. The grey lines are a power-law fitting model by decay slope, \( \alpha = 1.27 \pm 0.01 \)}
%    \label{fig:empirical_LC}
%\end{figure}

We start by analysing the afterglow of GRB~241030A empirically, that is using a power-law function of the form \( F_\nu \propto t^{-\alpha} \nu^{-\beta} \) to characterise its temporal and spectral evolution. We obtain a temporal decay slope \( \alpha = 1.23 \pm 0.01 \) and a spectral slope \( \beta = 0.70 \pm 0.01 \) across all X-ray and optical filter bands and before $\sim 10^4~\mathrm{s}$, when the X-ray light curve steepens slightly. These values are typical for long GRB afterglows (e.g. \citealt{2009MNRAS.397.1177E,2016ApJ...828...36D}) and in agreement with standard synchrotron emission from the region behind the forward shock \citep{1998ApJ...497L..17S}. 

\red{The long-term GRB afterglow synchrotron emission is usually expected to be in the slow-cooling regime for standard jet and environment parameters (e.g. \citealt{2002ApJ...568..820G}).} Assuming that all radiation is emitted behind the forward shock and that all measured bands fall between the minimum injection frequency $\nu_\mathrm{m}$ and the cooling frequency $\nu_\mathrm{c}$, we can use the analytical relations between \( p \), the electron energy distribution index, \( \alpha \), and \( \beta \) \citep{1998ApJ...497L..17S,2000ApJ...543...66P} to estimate \( p \) from the values of \( \alpha \). A homogeneous medium gives \( \alpha = \frac{3}{4}(p - 1) \), which results in \( p = 2.64 \pm 0.01 \), while a wind-like medium gives \( \alpha = \frac{3p - 1}{4} \), corresponding to \( p = 1.97 \pm 0.01 \). For the spectral slope, \( \beta = \frac{p - 1}{2} \) which yields \( p = 2.40 \pm 0.01 \). 
There is a discrepancy between the values of $p$ obtained from the spectral and the temporal analyses, which hints at a potentially more complex radiation scenario that includes other effects such as contributions from the lateral structure of the jet, or modifications to the emitted spectrum by SSC (synchrotron self-Compton) scatterings (see Section~\ref{sec:analytical_ssc}). The values of $p$ are substantially inconsistent in the case of a wind-like medium, while the value of \( p \sim 2.5 \) resulting from a homogeneous medium \citep{1998ApJ...497L..17S}, is a reasonable index for particle acceleration in shocks. In this scenario, the slight steepening of the X-ray light curve after $\sim 10^4~\mathrm{s}$ can be interpreted by the passage of $\nu_\mathrm{c}$ in the X-ray band.
%In this model, the relativistic ejecta interacts with the ambient medium, producing synchrotron radiation as it decelerates. The temporal and spectral evolution of the emission is governed by the microphysical parameters of the shock and the density profile of the surrounding environment. 
Based on this first analysis, we now investigate the scenario in which the observed afterglow is dominated by synchrotron radiation from a forward shock propagating in a homogeneous medium.

\subsection{Bayesian methods and data analysis}
\label{sec:models_presentation}

\begin{figure}
    \centering
    \includegraphics[width=\linewidth]{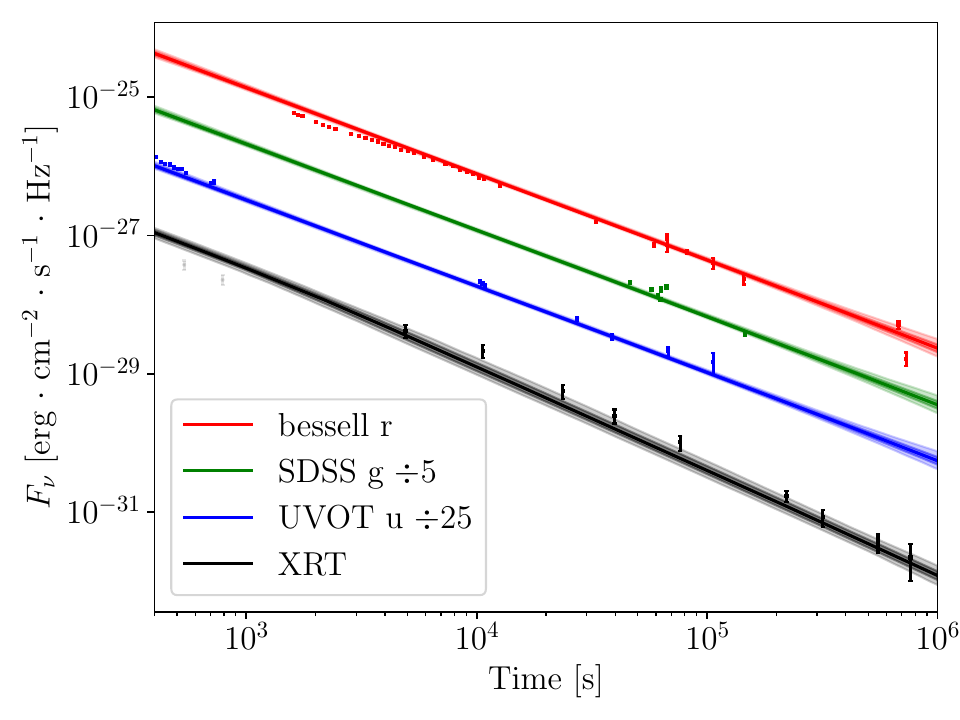}
    \caption{Light curve fits for selected filters from the NMMA inference with the \afterglowpy\ gaussian jet model.
   Observed data and their uncertainties are shown as error bars; the solid lines indicate the median value of the posterior sample at each time. 
   The shaded contours correspond to the 68\% and 95\% quantiles of the posterior distribution.
   The two transparent XRT data points before $10^3$~s were not included in the fit. %\rg{Isn't it a standard practice to present afterglow flux density in units of mJy?}
   }
    \label{fig:afgpy_lc}
\end{figure}

\begin{table}
    \centering
    \tabcolsep=0.25cm
    \caption{Parameters and priors used for the Bayesian analyses using the physical afterglow models. The two last columns contain the 95\% credible posterior interval for each particular parameter. The bar indicates that the parameter is not used by that model.}
    \begin{tabular}{>{\centering\arraybackslash}p{1.75 cm}
    >{\centering\arraybackslash}p{1 cm} >{\centering\arraybackslash}p{1.3 cm} >{\centering\arraybackslash}p{1. cm}>{\centering\arraybackslash}p{1.1 cm}}
\toprule 
Parameters & Symbol & Prior & \tiny{NMMA/} & PD24 \\ 
& & & \textsc{{\tiny afterglowpy}} &\\ 
\midrule 
cos inclination [rad] & $\cos(\iota)$ & $[0, 1]$ & $0.97^{+0.02}_{-0.03}$ & $0.99^{+0.01}_{-0.00}$ \\ 
log isotropic kinetic energy [erg] & $\log_{10}(E_0)$ & $[50, 60]$ & $56.9^{+1.6}_{-1.1}$ & $56.2^{+0.7}_{-0.7}$ \\ 
jet core angle [deg] & $\theta_{\text{c}}$ & $[0.056, 10]$ & $8.6^{+1.1}_{-2.3}$ & $9.7^{+0.1}_{-1.7}$ \\ 
wing factor & $\alpha_{\text{w}}$ & $[0.2, 3.5]$ & $2.72^{+0.72}_{-1.02}$ & - \\ 
log interstellar medium density [cm$^{-3}$] & $\log_{10}(n_{\text{ism}})$ & $[-6, 4]$  & $2.9^{+1.0}_{-4.6}$ & $2.5^{+0.7}_{-0.7}$ \\ 
electron spectrum power index & $p$ & $[2,3]$ & $2.62^{+0.05}_{-0.04}$ & $2.57^{+0.16}_{-0.06}$ \\ 
log electron energy fraction & $\log_{10}(\epsilon_\text{e})$ & $[-4, 0]$ & $-2.25^{+0.67}_{-1.27}$ & $-3.32^{+0.67}_{-0.66}$ \\ 
log magnetic energy fraction & $\log_{10}(\epsilon_\text{B})$ & $[-8, 0]$ & $-6.98^{+2.64}_{-0.90}$ & $-4.17^{+0.76}_{-0.72}$ \\ 
log accelerated electrons fraction & $\log_{10}(\zeta)$ & $[-3, 0]$ & - & $-0.70^{+0.67}_{-0.65}$ \\ 
initial Lorentz factor & $\Gamma_0$ & $[50, 500]$ & - & $432^{+65}_{-137}$ \\ 
systematic uncertainty [mag] & $\sigma_{\rm sys}$ & $[0.01, 2]$ & $0.21^{+0.03}_{-0.03}$ & - \\ 
\bottomrule
\end{tabular}
\label{tab:bayesian_priors}
\end{table}

To study the physical properties of the jet, we fit the observed light curve data with two physical GRB afterglow models.
Specifically, we rely on \afterglowpy~\citep{Ryan:2019fhz} and on the afterglow model from \cite{Pellouin:2024gqj}, hereafter referred to as PD24.
Both of these models assume that the emission arises when a laterally-structured relativistic jet interacts with the cool, constant-density interstellar medium and treat the forward shock dynamics and radiation emission in a semi-analytical fashion. The \afterglowpy~framework includes a treatment of the late-time lateral expansion of the jet, thus providing a more accurate calculation of the jet deceleration than PD24 in the trans-relativistic and Newtonian regimes. 
On the other hand, PD24 includes a treatment of synchrotron self-Compton (SSC) scattering in both Thomson and Klein-Nishina regimes, which can decrease the synchrotron emissivity and modify the synchrotron spectral shape (see \citealt{Pellouin:2024gqj, 2009ApJ...703..675N,Aguilar-Ruiz+25}). The initial coasting phase is also included in PD24.

In both analyses, we rely on the gaussian jet model where the initial isotropic-equivalent kinetic energy of the blastwave at any angle $\theta$ follows the distribution
\begin{align}
    E(\theta) = E_0 \exp\left(- \frac{\theta^2}{2\theta_c^2}\right)\ ,
\end{align}
with $\theta$ the polar coordinate angle measured from the jet axis and $\theta_c$ the core angle, a free parameter of both models, and $E_0$ the energy measured at the jet axis ($\theta=0$).
For the analysis with \afterglowpy\ we sample the posterior distribution of the light curve using the NMMA (Nuclear-physics and Multi-Messenger Astrophysics) Bayesian inference framework \citep{Pang:2022rzc}.
The posterior on the model parameters $\vec{\theta}$ is determined in NMMA by invoking the likelihood function
\begin{align}
\begin{split}
    \ln \mathcal{L}(\vec{\theta}|d) = - \sum_{t_j} \biggl(&\frac{1}{2} \frac{(m(t_j) - m^{\star}(t_j, \vec{\!\theta}\,))^2}{\sigma(t_j)^2 + \sigma_{\text{sys}}^2} \\
    &+ \ln(2\pi (\sigma(t_j)^2 + \sigma_{\text{sys}}^2)) \biggr).
    \label{eq:likelihood}
\end{split}
\end{align}
Here, $m(t_j)$ are the observed magnitudes with corresponding measurement uncertainties $\sigma(t_j)^2$, and $m^{\star}(t_j, \vec{\!\theta}\,)$ is the corresponding prediction of the \afterglowpy\ gaussian jet model for a given set of parameters $\vec{\theta}$. 
The parameter $\sigma_{\rm sys}$ accounts for possible systematic uncertainties caused by simplifying assumptions in the model.
Since the latter is hard to assess a priori, we introduce $\sigma_{\rm sys}$ as a sampling parameter and assess the need for a systematic error on-the-fly during the inference. 
This method has been suggested in \cite{Jhawar:2024ezm} and is implemented in NMMA.
Specifically, the prior for $\sigma_{\rm sys}$ is uniform between 0.01 and 2~mag.

%For the analysis with \textsc{PD24}, we focus on a top-hat jet model. \cp{CP: Modify this now that I used the gaussian model too.} We choose to ignore the lateral structure in this case given the monotonous light curve behaviour at all wavelengths. 
%The inclination is to let the core angle be a free parameter, which then converges to values smaller than the jet opening angle, therefore preventing significant contributions from a possible lateral structure. 
For the analysis with PD24, we use a similar gaussian jet model and the likelihood function is expressed as in Eq.~{\ref{eq:likelihood}}, without the introduction of the free parameter $\sigma_\mathrm{sys}$. 
Instead, we introduce a systematic contribution to the errors in the data such that each observation has at least a $0.2$~mag error. 
As shown by the fit with NMMA/\afterglowpy (see Fig. \ref{fig:nmma_corner}), this is a reasonable assumption, since $\sigma_\mathrm{sys}$ converges to values around that error.

The priors for the analyses are listed in Table~\ref{tab:bayesian_priors}. 
For the common parameters in the \afterglowpy~and PD24~models, the priors are kept identical. 
As discussed above, $\sigma_\mathrm{sys}$ is only introduced in NMMA/\afterglowpy, as well as the wing factor $\alpha_\mathrm{w}$ which determines the angular extent of the lateral structure. 
PD24 includes two additional free parameters: $\zeta$, describing the fraction of electrons which are accelerated behind the shock; and $\Gamma_0$, the initial Lorentz factor of the jet until the deceleration radius (during the coasting phase). Note that $\Gamma_0$ is kept uniform and does not have an angular profile.

We only fit data points measured more than $400~\mathrm{s}$ post-trigger time, to avoid any influence of the prompt emission on the analysis and focus on the external shock as suggested by \citet{2025ApJ...987..129W}. \red{The upper limits were not included in the fit, as none of them was expected to provide significant constraints, particularly given the large size of the dataset.}
Luminosity distance and redshift are fixed to $d_\mathrm{L}=10.3$\,Gpc and $z=1.411$, respectively.
Moreover, we exclude data points in the UV bandpasses that include wavelengths below the Lyman-$\alpha$ break for this particular redshift, i.e. below 2190~Å. 
Specifically, these are the \textit{Swift}-UVOT \textit{uvw1}, \textit{uvw2}, \textit{uvm2} and white filters.

In the analyses with NMMA/\afterglowpy, we exclude the LAT data points and the early XRT data before $10^3$~s.
%We estimated $10^3$~s as the end of the coasting phase from the \textsc{PD24} that characterise the early-time dynamics in the coasting phase where the Lorentz factor remains constant and estimate the deceleration radius of the material \citep{Pellouin:2024gqj}.
If we include the early XRT data in the NMMA/\afterglowpy\ analyses, it causes a bimodality in the inferred electron power law index $p$.
We suggest this behavior is due to the slight change of slope in the XRT data from early to late times, leading the \afterglowpy\ model to overestimate the flux at early times when fitting to the slope of the late-time X-ray data. 
To offset this, the sampler finds a second mode with higher $p$ that captures the early part of the light curve better.
However, this mode predicts even more extreme values for the jet energy and fraction of energy accelerating electrons $\epsilon_e$ as well as a larger inclination, and fits less well to the majority of the XRT data. Additionally, this set of solutions leads to coasting phase durations of about $10^3$~s, which are not accounted for by the model.
Hence, we conclude that this second mode is unphysical and exclude the early XRT data in the NMMA/\afterglowpy\ analyses to avoid impact on our conclusions. Observations between $400$ and $1000~\mathrm{s}$ are included in the fit with PD24.

\subsection{Results from NMMA/\textsc{afterglowpy} -- Synchrotron}
\label{sec:bayesian_syn}

The best-fit light curves for the \afterglowpy\ analysis are shown in Fig.~\ref{fig:afgpy_lc}. In Appendix~\ref{sec:ap_corner_plot}, we show the posterior distribution of the fits in Fig.~\ref{fig:nmma_corner} (light blue contours). We also show the posterior for a fit with PD24 that does not include SSC (pink contours), giving a very good consistency between the two models. In this section, we thus discuss only the results from NMMA/\afterglowpy. Overall, the model manages to fit the light curve within the expected systematic modeling uncertainties.
The fit slightly underestimates the early UV data, which could originate from an additional radiating region (e.g. a reverse shock, as discussed in \citealt{2025ApJ...987..129W}). 
%the lack of reverse shock treatment in the model, as also demonstrated in \citet{2025ApJ...987..129W}.
%Since \textsc{PD24} includes a coasting phase, it reproduces the early X-ray data well, which was excluded in the \afterglowpy~analyses.
The resulting posteriors (Appendix~\ref{sec:ap_corner_plot}, light blue contours) demonstrate that the NMMA/\afterglowpy\ sampling appropriately explores the parameter space and retrieves the degeneracies usually observed in afterglow modeling.

This analysis favors solutions corresponding to the analytical description of Section~\ref{section:empirical_fit}: the emission is powered by synchrotron emission of slow-cooling electrons between $\nu_\mathrm{m}$ and $\nu_\mathrm{c}$ in all bands until $\sim 10^4~\mathrm{s}$, where $\nu_\mathrm{c}$ crosses the X-ray band, which is a standard case of afterglow observations.

What deviates from standard afterglow theory is the values of some of the parameters inferred by the Bayesian fit. 
The analysis points to an energetic jet with $\log_{10}(E_0)~=~56.9_{-1.1}^{+1.6}$. (The intervals throughout this article mark the 95\% credibility limit.) The typical afterglow kinetic jet energy $E_0$ is usually limited to $\lesssim 10^{56} \mathrm{erg}$ \citep{Lloyd-Ronning:2004jxz, Aksulu:2021crt}, but the high energy we find for this GRB afterglow is in agreement with \citet{2025ApJ...987..129W}, as their analysis suggests $\log_{10}(E_0) = 55.83_{-0.18}^{+0.18}$. 
These high energy values are partially correlated to the low value for $\epsilon_e < 10^{-2}$ that our analysis also infers. The fraction $\epsilon_B$ is also found to be very low: $\epsilon_B \sim 10^{-7}$. As a secondary check, we thus reran the analyses with $\epsilon_e$ fixed to a more canonical value of $0.1$.
However, while these analyses indeed recover more moderate values for the energy at around $10^{55.5}$~erg, they deliver worse fits for most bands and increase the overall $\chi^2$-residuals notably. 
Fixing $\epsilon_e$ in the NMMA/\afterglowpy\ analyses also causes $n_{\rm ism}$ to rail against the upper prior bound of $10^4$ cm$^{-3}$ while the Bayesian evidence drops by a factor of 900 compared to the case where $\epsilon_e$ is left free. %These two checks does account for quantifying statistical and physical cost for testing a more low-energy solution in a more standard approach seen in other GRBs.

We investigated the energy budget of GRB~241030A by computing the kinetic energy corrected from the beaming, assuming a gaussian jet structure and using the results from our (most energetic) \texttt{afterglowpy} run (see Appendix~\ref{sec:ap_corner_plot}). We obtain 
\begin{align}
\begin{split}
    %E_\mathrm{kin} &= \int_0^{\theta_\mathrm{w}} d\theta\ E_0 \exp\left(- \frac{\theta^2}{2\theta_\mathrm{c}^2}\right)\ \sin^2(\theta/2) \\
    %&\approx 1.6 \times 10^{54}\ \rm{erg}\ .
    E_\mathrm{kin} &= 2\pi \int_0^{\theta_\mathrm{w}} d\theta\ \red{\sin(\theta)}\ \frac{E_0}{4\pi} \exp\left(- \frac{\theta^2}{2\theta_\mathrm{c}^2}\right)\ \\
    &\approx \red{8.7 \times 10^{54}\ \rm{erg}}\ .
\end{split}
\end{align}

In comparison, for the very bright GRB~221009A the estimates are between $ E_0 = 10^{54}$ and $10^{55}$ erg, but with a much narrower jet core, leading to more typical values of $E_\mathrm{kin}$ (e.g. \citealt{2023SciA....9I1405O, 2023MNRAS.524L..78G}). 
Under this scenario, this afterglow would be almost two orders of magnitude more energetic than GRB~221009A, making it one of the most energetic GRB afterglows so far. 
What stands out directly from our observations is that the afterglow of GRB~241030A is indeed very bright. We illustrate this by plotting the GRB~221009A $r'$ band light curve, placed at $z=1.411$, against our GRB~241030A light curve in Fig.~\ref{fig:grandma+community} (gray stars), for inter-comparison, which shows similar luminosities. However, the model parameters that we obtain here are much more extreme: given the GRB prompt duration of $300~\mathrm{s}$, accounting for breakout from the star, the jet power needed to accumulate this much energy corresponds to a jet luminosity of $L_\mathrm{j} \sim 5\times10^{53}\,{\rm erg\,s^{-1}}$, which may be difficult to explain in the collapsar scenario.

It is also noteworthy that, contrary to GRB~221009A, such a large kinetic energy is much larger than the prompt emission energy $E_\mathrm{iso} = 4 \times 10^{53}$~erg as determined in Section~\ref{section:E_iso_computation}, and our values for $E_0$ would imply a prompt emission efficiency of $\eta \lesssim 10^{-3}$, where we define $\eta = E_\mathrm{iso} / E_\mathrm{tot,iso} = 1 / (1 + E_0 / E_\mathrm{iso})$, where we assumed that the Lorentz factor of the ejected material along the line of sight is identical to the prompt Lorentz factor. %Note that one can argue that a more correct comparison would be to define $\eta$ using the kinetic energy at that angle, which is $E_{\rm kin,iso}(\theta_{\rm obs})10^{56}$\,erg for $\theta_{\rm obs}=14^\circ$ (consistent from Wang et al.,), as $E_{\rm iso}$ only radiates a small fraction of the total energy in the material within $1\Gamma_0$ around the LOS.
Typical expected values for $\eta$ are within $0.01$-- $0.9$~\citep{Beniamini:2016hzc, Racusin:2011jf}, though we should note that a new sample of low-efficiency long GRBs recently emerged due to more sensitive transient detection facilities~\citep{Sarin:2021eag, Ye2024, 2025ApJ...985..124L}.
We also performed a NMMA/\afterglowpy\ analysis where we set a lower limit on $\eta$ of 0.01, effectively constraining $\log_{10}(E_0) \leq 54.9$, however, this delivers a notably worse fit compared to a free $\log_{10}(E_0)$.
Specifically, we find a Bayes factor of $300,000$ in favor of the inference with a uniform prior $\log_{10}(E_0)\sim \mathcal{U}(50, 60)$ against the inference with a uniform prior $\eta \sim \mathcal{U}(0.01, 1)$. 
Additionally, we performed a fit with the \afterglowpy\ top-hat jet model, however this does not result in lower energy values either and is decisively disfavored with a Bayes factor of 230 compared to the gaussian jet model. Similar results are obtained with the top-hat model with PD24.
Finally, we also conducted a fit for a top-hat jet propagating in a wind environment with the PD24 model. This analysis did not perform better than for the case with a constant external density. 
We thus conclude that, if the afterglow is powered by synchrotron radiation of electrons accelerated behind a forward shock at the measured redshift $z=1.411$, the high jet energy coupled to a weak radiative efficiency (low $\epsilon_\mathrm{e}$, $\epsilon_\mathrm{B}$) offers the best explanation for the brightness observed in the afterglow of GRB~241030A. 

Moreover, our analyses find a relatively large opening angle for the core of the jet $\theta_\mathrm{c} = 8.6^{+1.1}_{-2.3}\degree$, which is impacted by our upper prior bound of $10\,\degree$. Such opening angles are higher than for typical GRB jets, but still consistent with jet ejection theory. 
%\rg{I'm not sure if there's a strong argument against have this large core angle.} This estimate is impacted by our upper prior bound of $10\,\degree$ for $\theta_{\rm c}$, and as such can be considered relatively high.
\cite{Goldstein:2015fib} typically find $\theta_{\rm c} < 15\,\degree$ for their sample of long GRBs. Moreover, the fact that we do not observe a visible jet break in our observations up to $\sim$8 days after the burst may indeed be  indicative of a large opening angle. While \citet{2025ApJ...987..129W} propose a jet break near $2\times10^5$~s and consequently find a smaller half-opening angle of $\theta_\mathrm{j}={4.5}^{+1.0}_{-0.7}\degree$, our observations are more complete at later times, in particular with the late \textit{EP} observation to confirm the X-ray behavior, and do not support this angular model and evolution.

\subsection{Analytical discussion}
\label{sec:analytical_ssc}

A striking result from the analysis presented in Section~\ref{sec:bayesian_syn} is the combination of a very high energy $E_0 \sim 10^{56}~\mathrm{erg}$ and a very low $\epsilon_\mathrm{B} < 10^{-6}$. Alternative parameter combinations with lower $E_0$ and higher $\epsilon_\mathrm{B}$ would yield comparable luminosities and are more commonly observed in GRB afterglows. We found that this unusual combination is favored by the sampler because of (1) the absence of a clear break in the optical light curve, meaning $\nu_\mathrm{m}$ should be below the optical bands at early times, and (2) to accommodate for the passage of $\nu_\mathrm{c}$ in the X-rays around $10^4~\mathrm{s}$. In the slow cooling regime, these two elements place strong constraints on the possible parameter space. We note from our results that the viewing angle is close to or within the jet core. We can therefore, with good accuracy, look at analytical formulae in the standard top-hat, on-axis case. Following \citet{2000ApJ...543...66P} and assuming that the emission is in the slow cooling regime, with $\nu_\mathrm{c}$ passing in the X-ray band at $10^4~\mathrm{s}$, we have the following four constraints:
\begin{itemize}
    \item[--] The model must reproduce the early time X-ray and optical temporal slope $\alpha$, which leads to a constraint on the injection slope $p$, $\alpha = \frac{3}{4} (p-1)$, for $\nu_\mathrm{m} < \nu < \nu_\mathrm{c}$ in a constant-density ISM;
    \item[--] The model must reproduce the normalization of the flux in the optical band, which gives a constraint on $E_0$, $n_\mathrm{ism}$, $\epsilon_\mathrm{e}$ and $\epsilon_\mathrm{B}$ (Eq.~B.7, 
    $F_\nu(t) \propto E_0^{(p+3)/4} n_\mathrm{ism}^{1/2} \epsilon_\mathrm{e}^{p-1} \epsilon_\mathrm{B}^{(p+1)/4} \nu^{-\beta} t^{-\alpha}$ 
    ). By construction, this constraint combined with the value for $p$ gives the correct normalization of the flux in X-rays before the break;
    \item[--] The passage of $\nu_\mathrm{c}$ in the X-rays at $10^4~\mathrm{s}$ gives a new constraint on $E_0$, $n_\mathrm{ism}$ and $\epsilon_\mathrm{B}$ (Eq.~27, 
    $\nu_\mathrm{c} \propto E_0^{-1/2} n_\mathrm{ism}^{-1} \epsilon_\mathrm{B}^{-3/2} t^{-1/2}$
    );
    \item[--] Because $\nu_\mathrm{m}$ only decreases with time, its value must be below the optical bands as early as the first data points considered, which gives another constraint on $E_0$, $\epsilon_\mathrm{e}$ and $\epsilon_\mathrm{B}$ (Eq.~22, 
    $\nu_\mathrm{m} \propto E_0^{1/2} \epsilon_\mathrm{e}^2 \epsilon_\mathrm{B}^{1/2} t^{-3/2}$
    ).
\end{itemize}
Starting from the best-fit parameters, working on the three last conditions and the corresponding equations implies the following correlations to any change of $E_0$, so that the normalization, and the values of $\nu_\mathrm{m}$ and $\nu_\mathrm{c}$ are kept identical:
\begin{equation}
\left\{
\begin{aligned}
\Delta \log(n_{\text{ism}}) &= -5 \, \Delta \log(E_0) \\
\Delta \log(\epsilon_\mathrm{e}) &= - \, \Delta \log(E_0) \\
\Delta \log(\epsilon_\mathrm{B}) &= 3 \, \Delta \log(E_0)
\end{aligned}
\right.
\end{equation}
Starting from $\log_{10} (E_0) = 56.6$ (best-fit model), a lower value of $\log_{10} (E_0) = 55$ maintaining these constraints would imply even more extreme values of $\log_{10} (n_\mathrm{ism}) = 10.6$, $\log_{10} (\epsilon_\mathrm{e}) = -1$ and $\log_{10} (\epsilon_\mathrm{B}) = -12$. Such values are even more extreme and well outside our range of priors.

With a very low magnetization ($\epsilon_\mathrm{B}\lesssim 10^{-6}$), it is expected that SSC should strongly affect the emitted spectrum throughout the afterglow. We employ the model of PD24 to explore this effect.

To illustrate the potential importance of SSC scatterings on the model predictions, we performed a diagnostic test by taking the results from the fits presented in Section~\ref{sec:bayesian_syn}, and computing the afterglow emission, this time accounting for SSC scatterings. A new fit, this time including SSC emission, is presented in Section~\ref{sec:bayesian_ssc}. %While this approach is not self-consistent, it serves to illustrate the potential impact of SSC on the predicted emission. 

\begin{figure}
    \centering
    \includegraphics[width=0.92\linewidth]{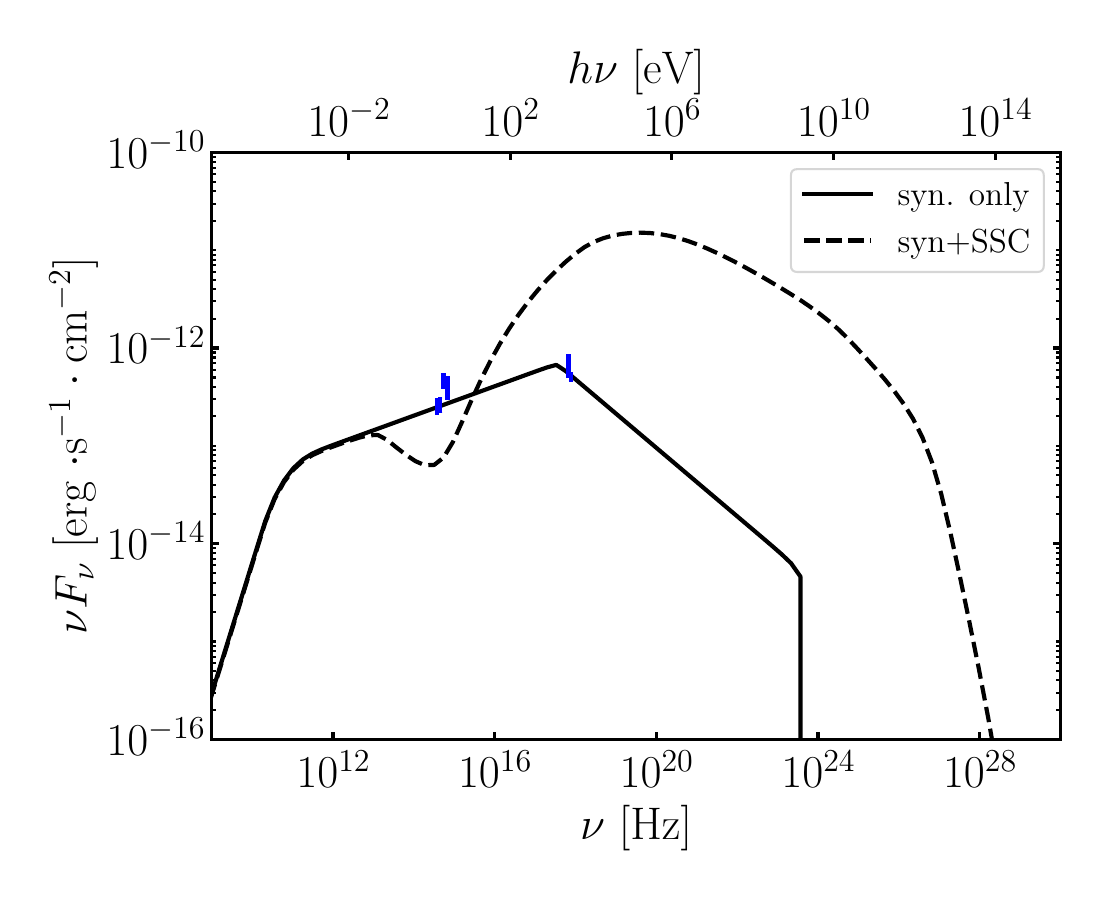}
    \caption{Observed spectrum at $t_\mathrm{obs} = 8 \times 10^4~\mathrm{s}$. The blue error bars show the fitted data points with $7.5\times10^4~\mathrm{s} < t_\mathrm{obs} < 8.5\times10^4~\mathrm{s}$ (optical and X-ray data). The solid curve represents the spectrum computed with the best-fit set of parameters found when SSC is not included in the model. The dashed curve represents the spectrum computed with the same parameters, this time including SSC.
    }
    \label{fig:sp_compare_SSC}
\end{figure}

All models corresponding to the highest likelihood posterior are significantly affected by SSC. The corresponding light curves then fail to match the afterglow observations. An example with the set of parameters which provide the best fit without SSC is provided in Fig.~\ref{fig:sp_compare_SSC}. This spectrum is taken after the passage of $\nu_\mathrm{c}$ in the X-ray band, which can be seen as the second break in the spectrum (solid line).

We can understand this effect by looking at the model parameters. It is noteworthy that, in the fits without SSC, $\epsilon_\mathrm{e}/\epsilon_\mathrm{B} > 10^{4}$ (see the light blue contours in Fig.~\ref{fig:nmma_corner}), which is an unusually high value and should lead to significant SSC emission. Indeed, if Klein-Nishina effects are ignored (SSC scatterings then occur only in the Thomson regime), then the ratio of SSC power to synchrotron power $Y_\mathrm{noKN} \sim \sqrt{\frac{\epsilon_\mathrm{e}}{\epsilon_\mathrm{B}}}$ in the fast cooling regime and $Y_\mathrm{noKN} \sim \sqrt{\frac{\epsilon_\mathrm{e}}{\epsilon_\mathrm{B}}} \left( \frac{\gamma_\mathrm{m}}{\gamma_\mathrm{c}^\mathrm{syn}}\right)^{\frac{p-2}{2}}$ in the slow cooling regime. In both cases, $Y_\mathrm{noKN} \gg 1$. Even with the inclusion of Klein-Nishina corrections in PD24, the synchrotron emission is significantly altered, as shown in Fig.~\ref{fig:sp_compare_SSC}. In this example, SSC scatterings occur in the Thomson regime, and $\nu_\mathrm{c}$ is significantly lowered below the optical band, while all the emission in the X-rays and at higher energies is dominated by SSC.

We therefore deduce from this analytical study that the highest likelihood posterior values in the case without SSC radiation are highly constrained by three elements that are difficult to reconcile: (i) a high optical and X-ray luminosity, driving high $E_0$ and $n_\mathrm{ism}$, (ii) $\nu_\mathrm{c}$ passing in the X-ray band at $\sim 10^4~\mathrm{s}$ while (iii) $\nu_\mathrm{m}$ is already below the optical bands at the earliest times (here $400~\mathrm{s}$), which strongly constrains the microphysics behind the shock. In addition, these sets of parameters (in particular, the high value of $\epsilon_\mathrm{e} / \epsilon_\mathrm{B}$) imply strong SSC emission. This affects the predictions and limits the validity of any model that ignores SSC scatterings (see Fig.~\ref{fig:sp_compare_SSC}). We therefore use the model of PD24 in Section~\ref{sec:bayesian_ssc} to assess other potential types of solution to the afterglow of GRB~241030A.

\subsection{Results from PD24 -- Synchrotron \& SSC}\label{sec:bayesian_ssc}

\begin{figure}
    \centering
    \includegraphics[width=0.9\linewidth]{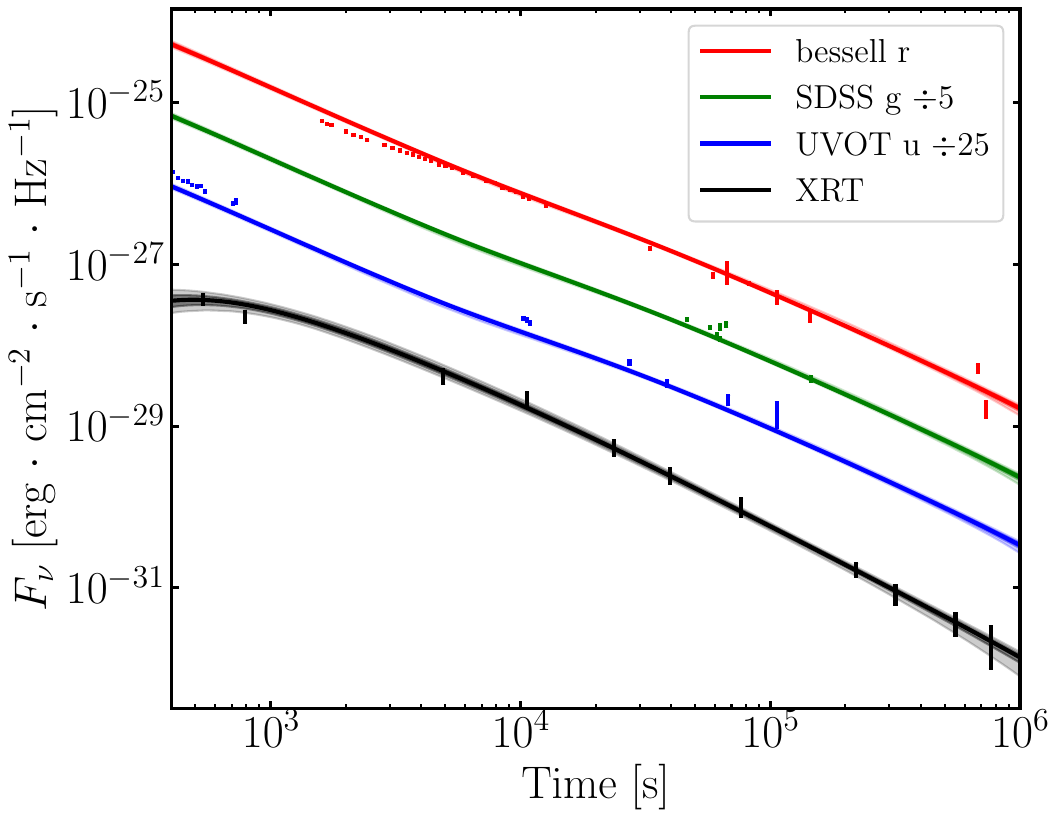}
    \caption{Light curve fits from the inference with the afterglow model from \citet{Pellouin:2024gqj} for selected filters.
    Observed data and their errors are shown, and the solid line indicates the median value of the posterior sample at each time. The shaded contours correspond to the $68\%$ and $95\%$ quantiles of the posterior distribution.
    }
    \label{fig:pellouin_lc}
\end{figure}

In this section, we show the results of the fit to the afterglow data from $400~\mathrm{s}$ post-$t_0$ with the PD24 model, including the effects of SSC and using the parameters and priors described in Section~\ref{sec:models_presentation} and Table~\ref{tab:bayesian_priors}. The best fit light curves for this analysis are shown in Fig.~\ref{fig:pellouin_lc}. The posterior distribution for this fit are shown in Fig.~\ref{fig:nmma_corner} (dark blue contours), and the $95\%$ credible intervals for each parameters are given in the last column of Table~\ref{tab:bayesian_priors}.

%A large kinetic energy and small opening angle raises the question about 
%\JGD{Comment on the energy budget with such large opening angle and large $E_{iso}$?}

With this analysis, we find a similar overall trend as found with NMMA/\afterglowpy: the jet is very energetic, $\log_{10}(E_0) = {56.2}^{+0.7}_{-0.7}$, and the jet opening angle is also large $\theta_c = 9.7^{+0.3}_{-1.7}\,\degree$, ramming against the upper limit of our prior. We again find very high external medium densities $\log_{10}(n_\mathrm{ism}) = {2.5}^{+0.7}_{-0.7}$. The analysis from PD24 suggests a lower value of $\epsilon_\mathrm{e} < 10^{-3}$, and higher values of $\log_{10} (\epsilon_\mathrm{B}) = {-4.17} ^{+0.76} _{-0.72}$. These differences are likely driven by the inclusion of SSC in the model, which affects in particular the value of $\nu_\mathrm{c}$.

The fitted data set includes two \textit{Fermi}-LAT observations. We found that our afterglow model under-predicts the emission in this band. This is a consequence of the larger errors on these observations and of the low statistical weight of these two points on the whole data set, which the sampling algorithm does not seek to optimize as much as the optical and X-ray observations.

\begin{figure}
    \centering
    \includegraphics[width=0.89\linewidth]{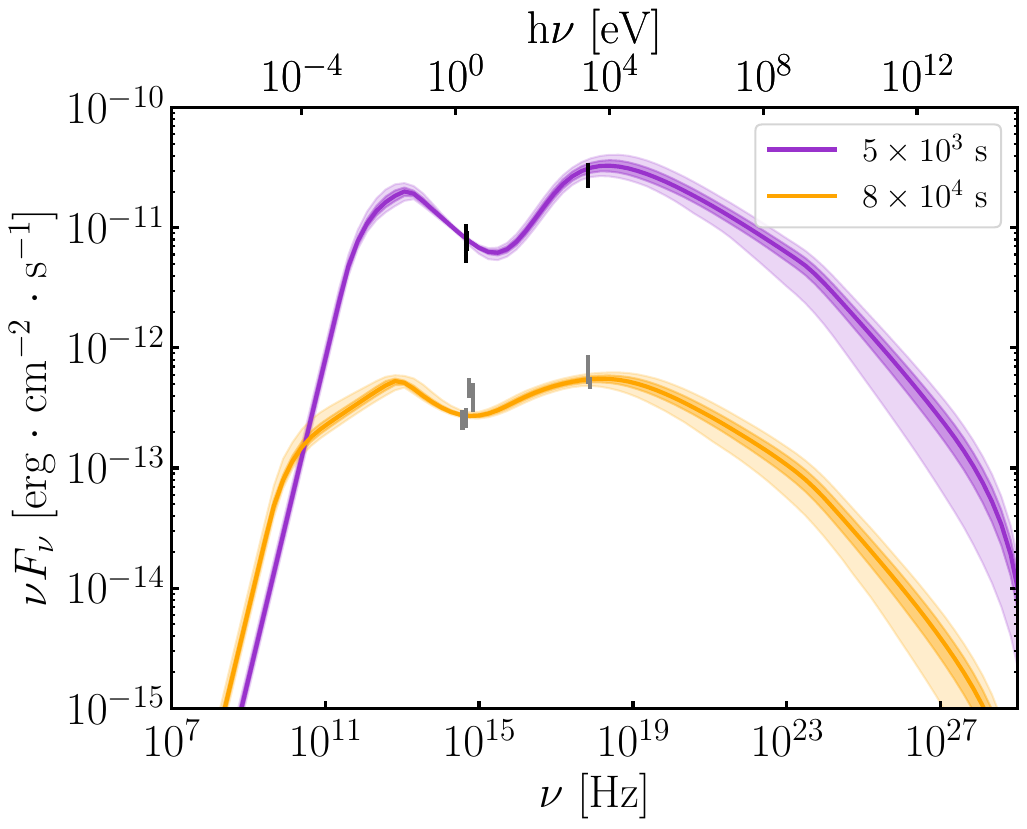}
    \caption{Afterglow spectra at $t_\mathrm{obs} = 5 \times 10^3~\mathrm{s}$ (purple) and $t_\mathrm{obs} = 8 \times 10^4~\mathrm{s}$ (orange) computed with the model from \citet{Pellouin:2024gqj}. The grey error bars show the fitted data points with $4\times10^3~\mathrm{s} < t_\mathrm{obs} < 6\times10^3~\mathrm{s}$, while the black error bars show the fitted data points with $7.5\times10^4~\mathrm{s} < t_\mathrm{obs} < 8.5\times10^4~\mathrm{s}$ (optical and X-ray data). The shaded contours correspond to the $68\%$ and $95\%$ quantiles of the posterior distribution.
    }
    \label{fig:pellouin_sp}
\end{figure}

We observe from Fig.~\ref{fig:pellouin_lc} that the multi-wavelength light curves are well reproduced by our analysis when SSC is included. However, as we show in Fig.~\ref{fig:pellouin_sp}, the X-ray LC is dominated by SSC emission, while the optical LC transitions through time from being synchrotron-dominated to SSC-dominated. This result is extremely unusual, and it is generally not expected that the X-ray emission be dominated by SSC. Overall, even the inclusion of SSC in the model does not resolve the tension detailed in Section~\ref{sec:analytical_ssc} and the sampling algorithm still favors extreme parameter values which provide a good fit to the data, while corresponding to a scenario for the radiating mechanisms beyond usual assumptions for afterglow models.

\section{Discussion}\label{sec:discussion}

%GRB~241030A was a bright long gamma-ray burst at redshift $z=1.411$ with a prompt emission detected in the gamma/X-ray range by several instruments. The rapid identification of a bright UV/optical afterglow counterpart led the community to conduct deep observations of its emission across the electromagnetic spectrum. This offers a rare well-sampled multi-wavelength afterglow light curve and an opportunity to constrain the GRB's jet structure, environment, and afterglow emission model.

%In this paper, we performed a detailed analysis of this gamma-ray burst's multi-wavelength afterglow, collecting gamma-ray, X-ray, UV, and optical data. 
%Because of this well-sampled multi-wavelength dataset, we constrained some of the main physical parameters of the standard GRB afterglow models. 

%The line-of-sight extinction analysis exhibits a Milky Way-like extinction with low $E(B-V)$. 
%To constrain some of the parameters of the forward shock model \citep{1998ApJ...497L..17S,Sari1998}, we performed both empirical and Bayesian fits of the light curves. Both provide a standard value for the electron distribution index $p$. 

GRB~241030A is a long GRB with a high-luminosity afterglow. We presented a complete multi-wavelength data set of the afterglow observations from optical to X-rays, and covering epochs up to $8 \times 10^5~\mathrm{s}$ ($\sim 9.3~\mathrm{days}$) post-trigger. In particular, we presented for the first time the observations from GRANDMA, KNC and COLIBRÍ. All optical/UV observations were processed in a consistent manner, and the consequences of line-of-sight extinction were coherently accounted for. Upon analysis of the optical images around the afterglow region, we did not find a host galaxy near the afterglow position, down to magnitudes $i<25.62$ and $r<25.82$. This suggests a relatively faint host galaxy.

We conducted multi-wavelength Bayesian analyses of the afterglow of GRB~241030A with two physical models: \afterglowpy\ \citep{Ryan:2019fhz} within the \textsc{NMMA} Bayesian framework \citep{Pang:2022rzc}, and the model from \citet{Pellouin:2024gqj}, which both assume radiation is emitted behind the forward shock of a structured jet, possibly viewed off-axis. \citet{Pellouin:2024gqj} also account for SSC scatterings. 
Our analyses conclusively show that this afterglow is powered by a highly energetic jet propagating in a high-density environment ($n_\mathrm{ism} \gtrsim 10^2~\mathrm{cm}^{-3}$). 
The absence of a clear multi-band jet break suggests that the emission is observed close to on-axis with a jet core angle ($\theta_c \gtrsim 8^\circ$), and that the jet edges have not yet become visible within the observed time span.

Beyond these general results, our analyses faced some challenges. 
The overall afterglow luminosity is very high and resulted in fits with high isotropic kinetic energy equivalents $E_0 \sim 10^{56}$~erg.
The absence of a jet break in the afterglow data further points to a large jet opening angle, our analyses find $\theta_c \approx 9\,\degree$ which is close to the upper boundary of our Bayesian prior.
These large estimates for the opening angles also lead to large values for the kinetic energy of the jet, of order $10^{54}~\mathrm{erg}$. Assuming that $1\%$ of the ejecta energy is converted into the jet, this would imply that the rest-mass equivalent energy of several tens of Solar masses be dynamically ejected before jet launching.

If our assumption that the emission is powered by a population of electrons emitting synchrotron and SSC photons behind a forward shock is correct, then our results have strong consequences for the physical conditions of this afterglow. The very high jet kinetic energy is in contrast with the prompt emission energy, which is a relatively standard $4\times 10^{53}$~erg, implying unusually low values for the prompt emission efficiency $\eta \lesssim 10^{-3}$.
Hence, GRB~241030A might be among a sample of a rare population of GRBs with very low prompt emission efficiency \citep{Sarin:2021eag, Ye2024, 2025ApJ...985..124L}. \citet{2025ApJ...985..124L} identified six afterglows discovered in optical-survey data between 2019 and 2023 with no detected prompt emission but with a measured redshift (between 1.06 and 2.9). \citet{Sarin:2021eag} had modeled AT2020blt (ZTF20aajnksq) and AT2021any (ZTF21aaeyldq), giving an efficiency $\lesssim$ 0.3–4.5\% for AT2020blt. \citet{Ye2024} studied AT2021lfa (ZTF21aayokph) and gave an "exceptionally low radiation efficiency" $\lesssim$ 0.02\%. \citet{2025ApJ...985..124L} added AT2023lcr (ZTF23aaoohpy) and gave an efficiency $\lesssim$ 2.3\%. Note in addition that the initial jet Lorentz factor has values $\Gamma_0 > 300$, strongly rejecting the hypothesis of being in the `dirty fireball' or `failed gamma-ray burst' situation \citep{2002MNRAS.332..735H, 2003ApJ...591.1097R}.
We also found unusually small values of $\epsilon_\mathrm{e}~(< 10^{-2})$ and $\epsilon_\mathrm{B}~(< 10^{-4})$, and fits where $\epsilon_\mathrm{e}$ was fixed to a more canonical value of $0.1$ were strongly disfavored. The combination of a very low magnetic energy fraction, a small electron energy fraction, and a large isotropic-equivalent kinetic energy suggests that the afterglow emission is largely adiabatic, with likely $\lesssim 1\%$ of the total energy being radiated.

This first analysis, assuming no SSC scatterings, gave quantitatively good fits, but corresponds to an extreme case of an otherwise quite standard scenario to explain the afterglow light curve. The optical bands are at all times in the $\nu_\mathrm{m} < \nu < \nu_\mathrm{c}$ regime, with a slow-cooling electron population. The X-rays are initially in the same regime and transition to $\nu > \nu_\mathrm{c}$ after $t\sim 10^4~\mathrm{s}$, leading to a mild steepening of the X-ray light curve. However, we found that the unusual physical parameters associated with the highest likelihood models lead in fact to a strong SSC contribution to the afterglow. When SSC is included, the resulting light curves and spectra with the same parameters fail to reproduce the observations.

We therefore relied on the model of \citet{Pellouin:2024gqj} to include SSC scatterings in the Thomson and Klein-Nishina regimes and conducted the same analysis. We found qualitatively similar results (a very energetic jet, seen within the jet core, with a large opening angle and unusually small values of $\epsilon_\mathrm{e}$ and $\epsilon_\mathrm{B}$). This time again, the multi-wavelength fit was satisfactory. However, in this scenario, we found a very strong SSC component, which is responsible for the emission in X-rays, and even down to the optical bands at late times. This scenario also challenges standard afterglow theory.

Analytical analysis in the synchrotron-only scenario suggested that these difficulties are due to the unusual combination of a very bright afterglow luminosity, the passage of $\nu_\mathrm{c}$ in the X-rays at $10^4~\mathrm{s}$, while simultaneously $\nu_\mathrm{m}$ remains below the optical bands even at the earliest times. These constraints impose unusually low values of $\epsilon_\mathrm{B}$, which naturally imply a strong SSC component ($Y_\mathrm{noKN} \propto (\epsilon_\mathrm{e} / \epsilon_\mathrm{B})^{1/2} (\gamma_\mathrm{m} / \gamma_\mathrm{c}^\mathrm{syn})^{(p-2)/2}$). When SSC is included, solutions favor X-ray emission from the SSC component, which releases the constraint on $\nu_\mathrm{c}$ but adds a new constraint to have very strong SSC emission. Therefore, the parameters differ slightly but the conclusions remain similar.

If the early-time optical afterglow emission is instead powered by another site (e.g. a reverse shock), then it is possible to release the constraint on $\nu_\mathrm{m}$ slightly. We performed another Bayesian fit with the model from \citet{Pellouin:2024gqj}, this time assuming that the optical afterglow only starts after $2\times10^4~\mathrm{s}$. We still found results with the same qualitative properties already discussed here, which suggests that our findings are intrinsic to the afterglow of GRB~241030A. In addition, in such a scenario another difficulty would be to explain the smooth connection without temporal or spectral break between the early and late optical light curves, if there are indeed two different emitting regions or processes.

Our analysed data set contains observations from optical to X-rays, with high energy photons up to 2 GeV . To better constrain the physical scenario behind this unusual afterglow, radio observations would be particularly useful, as they would provide constraints in the lower-frequency part of the spectrum below $\nu_\mathrm{m}$. This would therefore help constrain the value of $\nu_\mathrm{m}$ much better and thus clarify our conclusions. We also found that the \textit{Fermi}-LAT observations did not contribute significantly to constrain the fit, but in such a scenario with strong SSC radiation, very high energy (VHE) observations around $1~\mathrm{TeV}$ and by instruments such as the CTAO \citep{2019scta.book.....C} would provide very strong constraints on the SSC component and deepen our understanding of such afterglows.

\section{Conclusions}\label{sec:conclusion}

In summary, GRB~241030A exhibited a very bright GRB afterglow compared to the vast majority of the population. This work presents deep observations of its emission from gamma rays to IR. This kind of (bright afterglow) GRB offers a chance to obtain well-sampled multi-wavelength afterglow light curves and thus an opportunity to conduct sophisticated analyses to constrain the GRB jet structure, environment, and afterglow emission model.
We conducted both empirical and Bayesian analyses on the rich data of GRB~241030A which yielded extreme model parameters. They pointed to a powerful jet propagating in a dense medium, with isotropic-equivalent kinetic energies on the order of $10^{56}$ erg and large jet angles, supported by the absence of a jet break within 8~days. If these parameters can be trusted, they impose strong conclusions about a very low prompt gamma-ray emission efficiency $\eta \lesssim 10^{-3}$, even when synchrotron self-Compton (SSC) effects are accounted for. In this scenario, the emission in X-rays is dominated by SSC, which is also non-standard and challenges our model assumptions.
%The absence of a jet break in our analysis within 8~days supports the large-angle result, while the mild steepening observed in X-rays can be explained by the cooling break crossing the band. This afterglow still challenges our models. 
Our study shows that SSC scatterings play an important role in bright GRB afterglows and should be accounted for, but that even including this effect in our models may not be enough to reach a complete understanding of the radiative mechanisms in such events without complete spectral coverage of the afterglow observations from radio to very high energies. GRB~241030A might join the small but growing sample of long GRBs with low radiative efficiency.

\section*{Paper writing team contributions}

The global scientific research was equally led by J.-G.~Ducoin and C.~Pellouin. The Paper Writing Team (PWT) was coordinated by M.~Pillas and N.~Guessoum. Members of the PWT were (in alphabetical order): D.~Akl, S.~Antier, J.-G.~Ducoin, N.~Guessoum, A.~Holzmann Airasca, H.~Koehn, N.L.~Klinger, R. Z. ~Li,  Y.~Liang, F.~Magnani, J.~Mao, C.~Pellouin, R.~Pillera, N.~Rakotondrainibe, Z. Wang. The internal review was conducted by F.~Daigne, N.~Guessoum, J.~Mao, R. Gill. %Runs on the interpretation were performed by C.~Pellouin and H.~Koehn in particular.

\section*{Acknowledgments}
See details in the Appendix~\ref{Acknowledgements}.

\section*{Data availability}

Images and raw data are available upon request.

\bibliography{bibliography}{}
\bibliographystyle{aasjournal}

\appendix

\section{Additional Contributions}

\paragraph{Paper writing team} The global scientific research was equally led by J.-G.~Ducoin and C.~Pellouin. The Paper Writing Team was coordinated by M.~Pillas and N.~Guessoum. Members of the Paper Writing Team were in alphabetical order: D.~Akl, S.~Antier, J.-G.~Ducoin, N.~Guessoum, A.~Holzmann Airasca, H.~Koehn, N.L.~Klinger, R. Z. ~Li,  Y.~Liang, F.~Magnani, J.~Mao, C.~Pellouin, R.~Pillera, N.~Rakotondrainibe, Z. Wang. The internal review was conducted by F.~Daigne, N.~Guessoum, J.~Mao. Runs on the interpretation were performed by C.~Pellouin and H.~Koehn in particular.

\paragraph{Image processing} Data analysis of the images were performed by D.~Akl, S.~Antier, E.~de~Bruin, J.-G.~Ducoin, P.-A.~Duverne, S.~Karpov, N.~Kochiashvili, M.~Mašek, M.~Pillas, D.~Turpin. GRANDMA operations were carried out by C.~Douzet, P.-A.~Duverne, T.~Du~Łaź, T.~Hussenot, A.~Le~Calloch, M.~Mašek, N.~Kochiashvili, Y. Rajabov, M.~Tanasan, with assistance of the core-team: C.~Andrade, S.~Antier, M.~Coughlin, P.-A.~Duverne, P.~Hello, N.~Guessoum, S.~Karpov, T.~Pradier, D.~Turpin.

\paragraph{Observations}  Members of the AbAO team that contributed to observations are V.~Aivazyan, R.~Inasaridze, N.~Kochiashvili. Observations with AbAO-T150 were made by T.~Kvernadze. Members of the CFHT team that contributed to observations are S. Antier and M. Pillas. Members of the COLIBRÍ (outside of GRANDMA members) that contributed to observations are C.~Angulo, J.-L.~Atteia, S.~Basa, R.-L.~Becerra, D.~Dornic, F.~Fortin, L.~Garcia, R.~Gill, N.~Globus, W.~Lee, D.~Lopez-C.~Camara, E.~Moreno~Méndez, M.~Pereyra, D.~Ramírez, A.M.~Watson. Members of the FRAM team that contributed to observations are S.~Karpov, M.~Mašek. Members of the GTC team that contributed to observations are F.~Agui~Fernandez, A.~de~Ugarte~Postigo. Members of the KAO team that contributed to observations are G.~Hamed, E. Elhosseiny, R. Mabrouk, M.~Mohlam, A.~Takey. Members of the HAO team that contributed to observations are Z.~Benkhaldoun, A.~Kaeouach. Members of the NAO team that contributed to observations are B.M.~Mihov, L.~Slavcheva-Mihova. Members of the TAROT team that contributed to observations are S.~Antier, A.~Klotz, C.~Limonta. Members of the TRT team that contributed to observations are K.~Noysena, M.~Tanasan, K.S.~Tinyanont. 
Members of the TNOT that contributed to observations are A.~Esadmdin, A.~Iskandar, J.~Liu, H.~Peng, L.T.~Wang, X.F.~Wang, Y.S.~Yan. Members of the GMG that contributed to observations are J. Mao and R. Z. Li.

Members of the kilonova-catcher program of GRANDMA that contributed to observations are C.~Andrede, M.~Freeberg, H.B.~Eggenstein; R.~Kneip; A.~Lekic; M.~Odeh; M.~Serrau; D.~Turpin.

Members that contributed to the Einstein Probe team are Y.~Liang (Einstein Probe representative for this work), C.C.~Jin, H.~Sun, D.F.~Hu, S.Q.~Jiang, B.T.~Wang, H.Z.~Wu, Q.Y.~Wu, H.N.~Yang, X.F.~Wang, W.~Yuan, H.S.~Zhao, J.J.~Xu.

Members that contributed to the Fermi/LAT team are E.~Bissaldi; A.~Holzmann~Airasca; N.~Omodei; R.~Pillera.

Members that contributed to the Swift/BAT team are A.~Breeveld; P.A.~Evans; M.~Ferro coordinated by N.~Klingler.

\paragraph{Bayesian analysis} Runs on the interpretation were performed by C.~Pellouin and H.~Koehn in particular.

\section{Acknowledgments (details)}
\label{Acknowledgements}

M.W.~C. acknowledges support from the National Science Foundation with grant numbers PHY-2117997, PHY-2308862 and PHY-2409481. N.~G. acknowledges the support of the Simons Foundation (MP-SCMPS-00001470, N.~G.). H.~K. acknowledges funding from the European Union (ERC, SMArt, 101076369). N.~J.~K. acknowledges support by NASA under award number 80GSFC24M0006. J.~M. acknowledges support by the National Key R\&D Program of China (2023YFE0101200), Natural Science Foundation of China 12393813, and the Yunnan Revitalization Talent Support Program (YunLing Scholar Project). C.~P. acknowledges support by consolidator ERC grant 818899 (JetNS). A.M.~W. acknowledges support by UNAM/DGAPA project IN109224. The NRIAG team acknowledges financial support from the Egyptian Science, Technology \& Innovation Funding Authority (STDF) under grant number 45779.  B.M.~M. and L.~S.-M.  acknowledge support from the infrastructure purchased/renovated under the National Roadmap for Research Infrastructure (2020-2027), financially coordinated by the Ministry of Education and Science of Republic of Bulgaria (agreement D01-326/04.12.2023). M.~M. and S.~K. acknowledge support by MEYS (Czech Republic) under the projects Czech Republic MEYS LM2023032, LM2023047, and CZ.02.01.01/00/22\_008/0004632. The COLIBRÍ team acknowledges support of the staff of the Observatorio Astronómico Nacional on the Sierra de San Pedro Mártir and the COLIBRÍ and DDRAGO engineering teams.

AbAO team acknowledges Shota Rustaveli National Science Foundation of Georgia (SRNSFG, grant FR-24-7713). E.~B., A.~H.~A., N.~O. and R.~P. acknowledge support for LAT development, operation and data analysis from NASA and DOE (United States), CEA/Irfu and IN2P3/CNRS (France), ASI and INFN (Italy), MEXT, KEK, and JAXA (Japan), and the K.A. Wallenberg Foundation, the Swedish Research Council and the National Space Board (Sweden). This work is also supported in the operations phase from INAF (Italy), and CNES (France) is also gratefully acknowledged. This work was performed in part under the DOE Contract DEAC02-76SF00515. This work is partly based on data obtained with the CFH/Megacam instrument, and the ACME is thanked for providing necessary time for this work. COLIBRÍ received support from the French government under the France 2030 investment plan, as part of the Initiative d’Excellence d’Aix-Marseille Université-A*MIDEX through (ANR-11-LABX-0060 - OCEVU) and (AMX-19-IET-008 - IPhU), from LabEx FOCUS (ANR-11-LABX-0013), From Centre National d'Etudes Spatiale (CNES) and from CSAA-INSU-CNRS support program, and in Mexico from UNAM (Secretaria Administrativa, Coordinacion de la Investigacion Cientıfica, Instituto de Astronomıa and PAPIIT grant IN105921), and SECIHTI/CONACyT (277901, Ciencias de Frontera 1046632 and Laboratorios Nacionales). This work is also partly based on data obtained with the instrument OSIRIS, built by a Consortium led by the Instituto de Astrof\'isica de Canarias in collaboration with the Instituto de Astronom\'ia of the Universidad Aut\'onoma de México. OSIRIS was funded by GRANTECAN and the National Plan of Astronomy and Astrophysics of the Spanish Government. This work is finally based on observations made with the Thai Robotic Telescope under program ID TRTC12A\_001, which is operated by the National Astronomical Research Institute of Thailand (Public Organization). CFH and Skyportal are part of a project that has received funding from the European Union’s Horizon Europe Research and innovation programme under Grant Agreement No 101131928. This project is based on observations obtained with MegaPrime/MegaCam, a joint project of CFHT and CEA/DAPNIA, at the Canada-France-Hawai'i Telescope (CFHT) which is operated by the National Research Council (NRC) of Canada, the Institut National des Sciences de l'Univers of the Centre National de la Recherche Scientifique (CNRS) of France, and the University of Hawai'i. C.~D. acknowledges funding from the European Union’s Horizon Europe Research and Innovation Programme under Grant Agreement No 101131928 (ACME project).

% FERMI-LAT acknowledgement paragraph
The \textit{Fermi} LAT Collaboration acknowledges generous ongoing support from a number of agencies and institutes that have supported both the development and the operation of the LAT as well as scientific data analysis. These include the National Aeronautics and Space Administration and the Department of Energy in the United States, the Commissariat \`a l'Energie Atomique and the Centre National de la Recherche Scientifique / Institut National de Physique Nucl\'eaire et de Physique des Particules in France, the Agenzia Spaziale Italiana and the Istituto Nazionale di Fisica Nucleare in Italy, the Ministry of Education, Culture, Sports, Science and Technology (MEXT), High Energy Accelerator Research Organization (KEK) and Japan Aerospace Exploration Agency (JAXA) in Japan, and the K.~A.~Wallenberg Foundation, the Swedish Research Council and the Swedish National Space Board in Sweden.

Additional support for science analysis during the operations phase is gratefully  acknowledged from the Istituto Nazionale di Astrofisica in Italy and the Centre  National d'\'Etudes Spatiales in France. This work performed in part under DOE  Contract DE-AC02-76SF00515.

%\section{Swift-XRT data analysis}
%\label{sec:afterglowXray}

%We conduct a single analysis on the \textit{Swift}-XRT X-ray afterglow fading epoch. It is best described by a power-law with two breaks using the \href{www.swift.ac.uk/xrt_live_cat/01263718}{XRT Live catalog} \citep{2007A&A...469..379E,2009MNRAS.397.1177E}: $\alpha_1 = 1.44 \pm 0.03$, $T_{\rm break,1} = 689^{+69}_{-75}$ s, $\alpha_2 = 0.94^{+0.03}_{-0.04}$, $T_{\rm break,2} = 1010^{+134}_{-185}$ s, and $\alpha_3 = 1.52\pm0.05$.  The results are compatible with the multi-band results mentioned in Section~\ref{section:empirical_fit}.

\section{Details about Fermi-LAT data reduction}
\label{sec:lat_appendix}

For the LAT analysis we followed the standard approach (as described in \citealt{2cat}) using the ``P8R3`` \citep{pass8,bruel2018fermilatimprovedpass8event} processed events. Specifically, we employed the \texttt{P8R3\_TRANSIENT\_010E} event class, during the time interval where we have significant detections, which is optimized for transient source analysis, along with the corresponding instrument response functions\footnote{\url{http://www.slac.stanford.edu/exp/glast/groups/canda/lat Performance.htm}}. For the extended emission search, we used the cleaner class \texttt{P8R3\_SOURCE}.

Events were chosen within the time interval from $t_0$ to $t_0 + 13 \mathrm{ks}$ covering the energy range from 100 MeV to 100 GeV in a region of $10^{\circ}$ centered on the UVOT position. For the spectral analysis we used ``Multi-Mission Maximum Likelihood`` (\texttt{ThreeML}) \citep{3ML}, a python-based software package for parameter inference with maximum likelihood, that provides an interface to the instrument response files and data in plugins. For the LAT data, two different plugins are available: \texttt{FermiLATLike} for unbinned analysis (which interfaces with \texttt{gtburst}\footnote{\url{https://fermi.gsfc.nasa.gov/ssc/data/analysis/scitools/gtburst.html}}, part of the \texttt{fermitools} \footnote{\url{https://github.com/fermi-lat/Fermitools-conda/}}), and \texttt{FermipyLike} for binned likelihood analysis with and interface to \texttt{fermipy} \citep{fermipy_wood}. The results of the time resolved spectral analysis are shown in Table~\ref{tab:lat_data}.

\begin{figure}
    \centering
    \includegraphics[width=.5\textwidth]{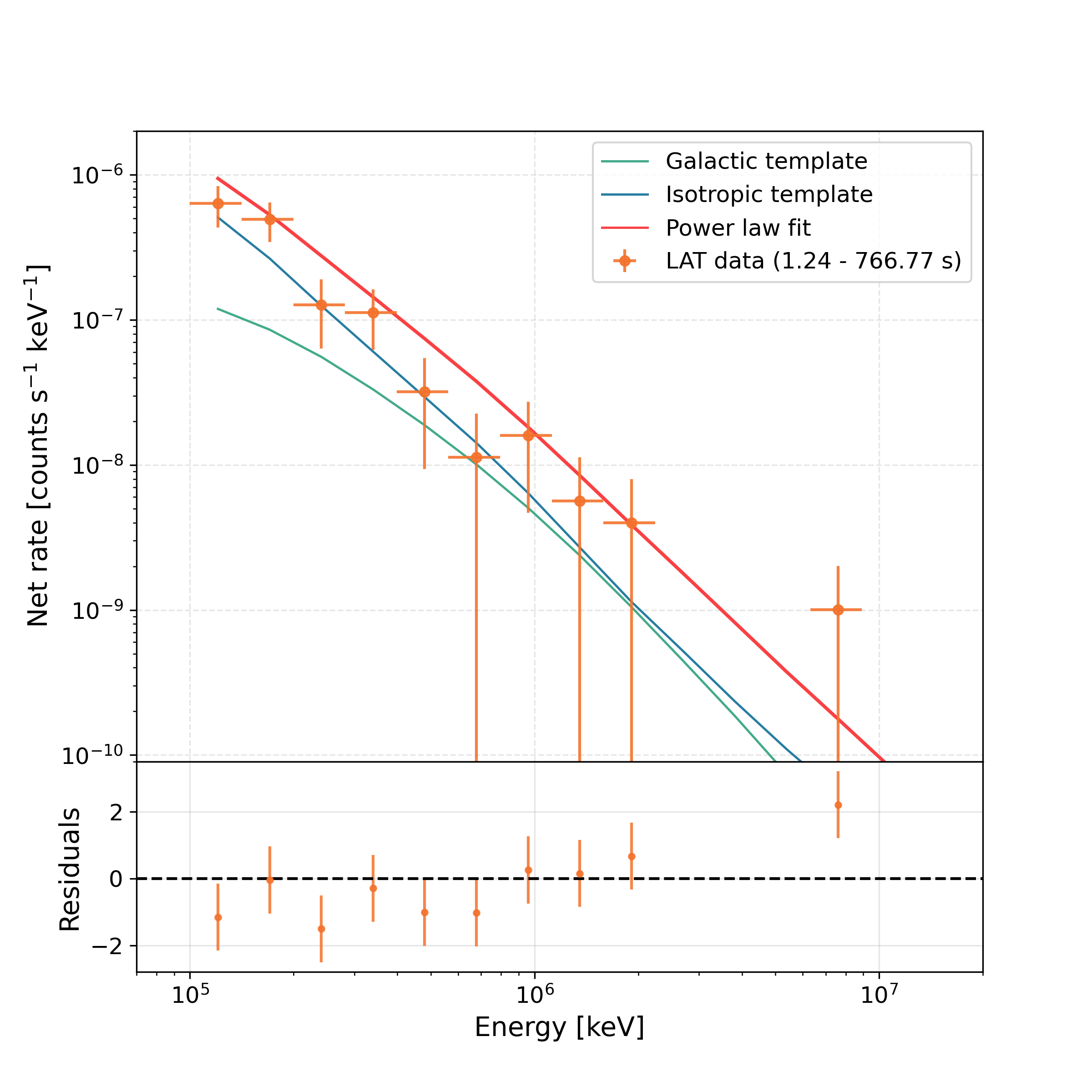}
    \caption{Fermi-LAT count spectrum (in counts s$^{-1}$ keV$^{-1}$) and residuals (in sigma units) for the \texttt{PL} model for the time integrated analysis.}
    \label{fig:lat_time_integrated}
\end{figure}

We performed an unbinned likelihood analysis of the burst spanning from $t_0$ to $t_0 + 800 \mathrm{s}$, using the \texttt{FermiLATLike} plugin, using the \texttt{P8R3\_TRANSIENT\_010E} event class. This analysis returned a test-statistic value of TS $\sim 49$, corresponding to a significance $\sigma \sim 7$, using $\sigma \sim \sqrt{\text{TS}}$. The result of the spectral analysis is displayed in Fig.~\ref{fig:lat_time_integrated}.

\setlength{\tabcolsep}{3pt}
\begin{table}[h!]
\centering
\begin{tabular}{cccccr}
\toprule
Interval & Flux \scriptsize{(MeV s$^{-1}$ cm$^{-2}$)} & Spec. index  & TS & Class\\
$T - T_{0}$ [s] & 0.1 - 10 GeV &  $\Gamma$ &  & \\
\midrule
1.2 - 292 & $(1.0^{+1.2}_{-0.5})\times10^{-3}$ & $-2.60^{+0.78}_{-0.82}$ & 11.3 & T\\
292 - 403 & $(4.2^{+4.9}_{-2.0})\times10^{-3}$ & $-1.91^{+0.49}_{-0.48}$ & 10.4 & T\\
403 - 556 & $(2.8^{+3.2}_{-1.3})\times10^{-3}$ & $2.0^{+0.46}_{-0.46}$ & 22.8 & T\\
556 - 767 & $(3.3^{+2.4}_{-1.2})\times10^{-3}$ & $-2.03^{+0.38}_{-0.37}$ & 21.8 & T\\
5.3k - 7.1k & $<1.7\times10^{-3}$ & $-2.50$  & 1.18 & S \\
9.7k - 13k & $<5.2\times10^{-4}$ & $-2.50$  & 0.39 & S \\
15.5k - 22.7k$^*$ & $<1.1\times10^{-5}$ & $-2.50$  & -0.34 & S \\
22.7k - 33.1k$^*$ & $<1.2\times10^{-5}$ & $-2.50$  & -0.25 & S \\
33.1k - 48.4k$^*$ & $<1.1\times10^{-5}$ & $-2.50$  & -0.57 & S \\
48.4k - 70.6k$^*$ & $<1.0\times10^{-5}$ & $-2.50$  & -0.50 & S \\
70.6k - 103.1k$^*$ & $<4.0\times10^{-4}$ & $-2.50$  & 0.82 & S \\
103.1k - 150.6k$^*$ & $<6.0\times10^{-4}$ & $-2.50$  & 0.34 & S \\
150.6k - 219.9k$^*$ & $<1.0\times10^{-5}$ & $-2.50$  & -0.73 & S \\
219.9k - 321.2k$^*$ & $<1.0\times10^{-5}$ & $-2.50$  & -1.07 & S \\
321.2k - 469.0k$^*$ & $<9.9\times10^{-5}$ & $-2.50$  & 1.98 & S \\
469.0k - 684.8k$^*$ & $<1.1\times10^{-5}$ & $-2.50$  & -2.21 & S \\
\bottomrule
\end{tabular}
\caption{Time-resolved spectral analysis of GRB\,241030A with LAT. The symbol $^*$ indicates a binned analysis (unbinned otherwise). A detection is considered significant in case of TS$>9$. For upper limits, the spectral index has been fixed to $-2.50$. The event class "T" refers to \texttt{P8R3\_TRANSIENT\_010E}, and "S" refers to \texttt{P8R3\_SOURCE}.}
\label{tab:lat_data}
\end{table}

%\section{Spectroscopic observation}\label{ap:spectrum}

%\begin{figure*}[htbp]
%  \centering
%  \includegraphics[width=\linewidth]{figures/GRB241030A_GMG_spec}
%  \caption{Optical spectrum of GRB~241030A obtained with the GMG 2.4-meter telescope. The wavelength-calibrated 2D spectrum and the flux-calibrated 1D spectrum are displayed. The 1D raw, smoothed, and uncertainty spectra are plotted in sky blue, blue, and gray lines, respectively. Several metal absorption lines at a redshift of $z = 1.411$ are indicated with vertical dotted lines.}
%  \label{fig:GRB241030A_GMG_spec}
%\end{figure*}

\section{Posterior corner plot for the Bayesian analyses}\label{sec:ap_corner_plot}

\begin{figure*}
    \centering
    \includegraphics[width=1.0\linewidth]{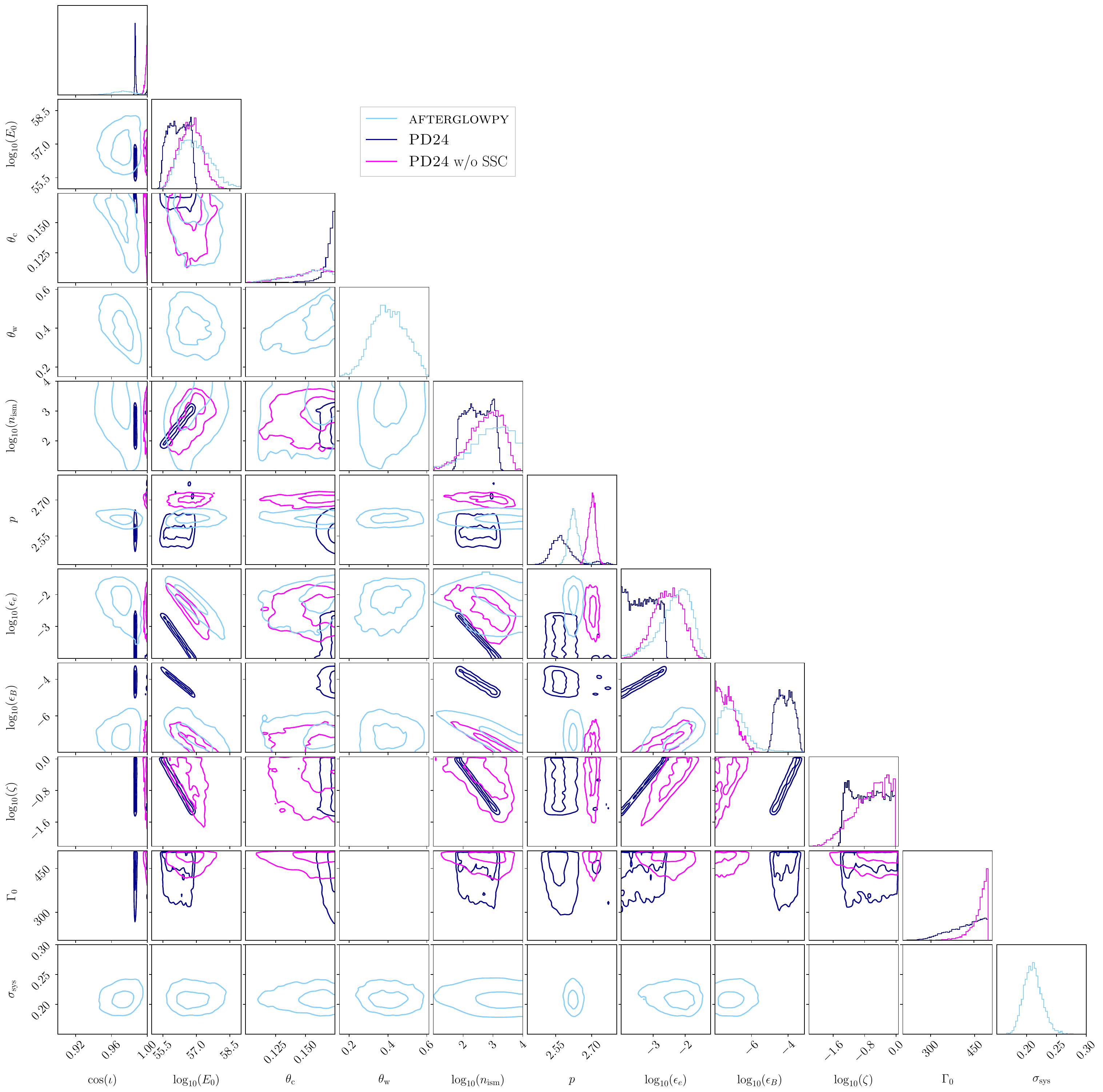}
    \caption{Posterior corner plot for the Bayesian analyses with the NMMA/\afterglowpy\ gaussian jet model (light blue) and the PD24 model without SSC (pink) and with SSC (dark blue).
    Due to the different model parameters, certain columns and rows only show the posterior for either the NMMA/\afterglowpy\ inference or the inference with PD24.
    The contours show the 39\% and 86\% credible regions.}
    \label{fig:nmma_corner}
\end{figure*}

We show the posterior corner plots from the inferences in Section~\ref{section:MWafterglowanalysis} in Fig.~\ref{fig:nmma_corner}.

\section{Observational campaign}\label{Appendix:campaign}

GRB~241030A was detected at $t_0$ = 05:48:03 UT on 2024 October 30 ($T_{90}=166$\,s in 50--300 keV) by the \textit{Fermi} Gamma-ray Burst Monitor (GBM, \citealt{2024GCN.38015....1D}). It was also independently detected by \textit{Konus-Wind} (\textit{KW}, \citealt{2024GCN.37982....1R}) at 2024-10-30 05:49:56 in the 18~keV--1~MeV band ($T_{90}=211$\,s), by \textit{Gamma-Ray Burst Alpha} (\textit{GRBAlpha}, \citealt{GCN38074}) at 2024-10-30 05:50 in the 80--950 keV band ($T_{90}=208$\,s), by the \textit{SVOM} Gamma-Ray Monitor (GRM) at 2024-10-30 05:48:14 in the 15~keV--5~MeV range \citep{GCN37972}, and by the \textit{Neil Gehrels Swift Observatory (Swift)} Burst Alert Telescope (BAT, \citealt{2024GCN.37956....1K}) in the 15--150~keV band ($T_{90}=173$\,s, \citealt{GCN38010}). Further characteristics of its prompt emission are described in \citet{2025ApJ...987..129W,2025arXiv251024864V}.

The optical counterpart was rapidly detected by the \textit{Swift} Ultraviolet/Optical Telescope (UVOT), providing an accurate position of the event (RA = 343.13987, DEC = 80.44996, 0.61 arcsec uncertainty, \citealt{2024GCN.37956....1K}) thanks to \textit{Swift}'s rapid slewing capabilities \citep{2005SSRv..120...95R}. As mentioned in \citet{2024GCN.38055....1W}, the UVOT instrument covered part of the UV and optical prompt emission as observations started 80s post-$t_0$. This prompt sequence was followed by a multi-wavelength campaign in the hard X-ray, soft X-ray, optical, and infrared bands, including: the \textit{Transiting Exoplanet Survey Satellite} (\textit{TESS}, \citealt{2024GCN.38134....1J}), the \textit{COSMOS-8k} camera onboard the 1-m telescope at Palomar \citep{2024GCN.38021....1S}, the Special Astrophysical Observatory of the Russian Academy of Sciences (SAO RAS, \citealt{2024GCN.38016....1M}), and the \textit{SVOM} Chinese Ground Follow-up Telescope (C-GFT, \citealt{2024GCN.37970....1S}). 

In this work, we collected data from many facilities. First, the GRB afterglow was detected at high energies by the \textit{Fermi} Large Area Telescope (LAT, \citealt{LAT_GCN}). In X-rays, the Wide-field X-ray Telescope (WXT, \citealt{2024GCN.37997....1W}) and the Follow-up X-ray Telescope (FXT, \citealt{2024GCN.38026....1L}) onboard the \textit{Einstein Probe} (\textit{EP}, \citealt{Yuan_2022}), together with the \textit{Swift}-XRT, also detected the X-ray afterglow \citep{2024GCN.37956....1K}. On the ground, we can cite in particular the Global Rapid Advanced Network for Multi-messenger Addicts (GRANDMA; \citealt{10.1093/mnras/staa1846}; \citealt{2022MNRAS.515.6007A}; \citealt{2023ApJ...948L..12K}; \citealt{2024A&A...682A.141T}) network, which includes contributions from the Abastumani Observatory 70-cm telescope and 1.5m telescope (AbAO-T70, AbAO-T150), the MegaCam instrument onboard the Canada–France–Hawaii Telescope (CFHT), the F/(Ph)otometric Robotic Atmospheric Monitor (FRAM), the Gran Telescopio Canarias with the Optical System for Imaging and low-Intermediate-Resolution Integrated Spectroscopy (GTC/OSIRIS), the 50-cm telescope of the High Atlas Observatory (HAO), the 2-m telescope of the Kottamia Astronomical Observatory (KAO), the 1.5-m AZ1500 and 2-m telescopes of the Rozhen National Astronomical Observatory of Bulgaria (NAO-1.5m, NAO-2m), the Thai Robotic Telescope at Sierra Remote Observatories (TRT/SRO), the Telescope à Actions Rapides of Calern (TAROT/TCA), and the 80-cm Tsinghua-Nanshan Optical Telescope (TNOT).  This network during the follow-up was also associated with the COLIBRÍ telescope, equipped with the commissioning camera named OGSE, and the GRANDMA/Kilonova-Catcher (KNC) amateur program\footnote{\raggedright{\url{http://kilonovacatcher.in2p3.fr/}}}.
In addition to the photometric reports, a spectroscopic redshift measurement of GRB~241030A yields $\mathrm{z}=1.411$ \citep{Zheng2024_GCN37959, 2024GCN.38027....1L}. Similarly (13:15:33 UT on 2024 October 30), we also conducted spectroscopic observations of GRB~241030A during 1h using the Gao-Mei-Gu (GMG) 2.4-meter telescope at Lijiang Observatory \citep{2020RAA....20..149X}. The spectroscopic observations used the GMG grating–prism disperser \#3, which provides a wavelength coverage of 3400–9100 \AA\ \citep{2019RAA....19..149W}, but we were not able to identify a clear signature allowing redshift determination.

\section{Observational data}\label{ap:data}

\begin{table}[t]
\centering
\normalsize
\setlength{\tabcolsep}{1.5pt}
\caption{Gamma/X-ray observations of GRB~241030A. Column (1) is the time delay 
between the observation and the GRB $T_{0}$ in days. Column (3) is the 
unabsorbed flux density in Jy. Flux densities are computed at 
$2.77~\mathrm{keV}$ for the \textit{Swift}-XRT, \textit{EP}-WXT and 
\textit{EP}-FXT instruments, and at $1~\mathrm{GeV}$ for \textit{Fermi}-LAT. 
In Column (5), X indicates data points not used in the Bayesian light-curve 
analysis, while A and P denote data used in NMMA/\afterglowpy\ or PD24, 
respectively.}
\label{tab:grandmadata_he}

\begin{tabular}{cccccc}
\toprule
$\Delta T$ & MJD & Flux & Instrument & Analysis \\
(days) &  & (Jy) &  &  \\
\midrule
0.00116 & 60613.24286  & $1.13 \pm 0.153 \times 10^{-3}$ & Swift-XRT & X \\
0.00170 & 60613.243401 & $1.07^{+1.28}_{-0.53} \times 10^{-10}$ & Fermi-LAT & X \\
0.00194 & 60613.243645 & $2.63 \pm 0.348 \times 10^{-3}$ & Swift-XRT & X \\
0.00315 & 60613.244847 & $1.02 \pm 0.169 \times 10^{-3}$ & Swift-XRT & X \\
0.00403 & 60613.245727 & $6.00^{+7.00}_{-2.86} \times 10^{-10}$ & Fermi-LAT & X \\
0.00428 & 60613.245983 & $1.88 \pm 0.257 \times 10^{-4}$ & Swift-XRT & X \\
0.00555 & 60613.247253 & $4.03^{+4.60}_{-1.87} \times 10^{-10}$ & Fermi-LAT & P \\
0.00622 & 60613.247925 & $3.77 \pm 0.575 \times 10^{-5}$ & Swift-XRT & P \\
0.00766 & 60613.249357 & $4.74^{+3.45}_{-1.73} \times 10^{-10}$ & Fermi-LAT & P \\
0.00914 & 60613.250838 & $2.28 \pm 0.375 \times 10^{-5}$ & Swift-XRT & P \\
0.03142 & 60613.273125 & $1.17 \pm 0.636 \times 10^{-5}$ & EP-WXT & A--P \\
0.05687 & 60613.298567 & $4.20 \pm 0.920 \times 10^{-6}$ & Swift-XRT & A--P \\
0.12295 & 60613.364652 & $2.16 \pm 0.478 \times 10^{-6}$ & Swift-XRT & A--P \\
0.27378 & 60613.515478 & $5.57 \pm 1.23 \times 10^{-7}$ & Swift-XRT & A--P \\
0.45894 & 60613.700644 & $2.47 \pm 0.563 \times 10^{-7}$ & Swift-XRT & A--P \\
0.88551 & 60614.127212 & $1.02 \pm 0.256 \times 10^{-7}$ & Swift-XRT & A--P \\
0.90176 & 60614.143461 & $6.66 \pm 0.653 \times 10^{-8}$ & EP-FXT & A--P \\
2.55774 & 60615.799437 & $1.70 \pm 0.321 \times 10^{-8}$ & Swift-XRT & A--P \\
2.84056 & 60616.082257 & $1.14 \pm 0.266 \times 10^{-8}$ & EP-FXT & A--P \\
3.67543 & 60616.917135 & $8.39 \pm 2.33 \times 10^{-9}$ & Swift-XRT & A--P \\
4.45853 & 60617.700231 & $5.53 \pm 1.31 \times 10^{-9}$ & EP-FXT & A--P \\
6.38300 & 60619.624706 & $3.72 \pm 1.15 \times 10^{-9}$ & Swift-XRT & A--P \\
8.84158 & 60622.083285 & $2.20 \pm 1.20 \times 10^{-9}$ & Swift-XRT & A--P \\
\bottomrule
\end{tabular}
\end{table}

%\twocolumn

%\input{data_table_GRB241030a_HE}

%\onecolumn
\newpage
\setlength{\tabcolsep}{1pt}
\begin{table}[]
\centering
\caption{UV/Optical observations of GRB~241030A. Column (1) is the time delay between the observation and the GRB $T_{0}$ in days. Column (3) gives apparent magnitudes in the AB system, without any correction. In Column (6), a X means we did not use this data point for the Bayesian light curve analysis, an A or a P means this data point has been used for NMMA/\afterglowpy or PD24, respectively.} 
\begin{tabular}{cccccr}
\toprule
\hline
\boldmath $\Delta T$ \unboldmath & \textbf{MJD} & \textbf{Magnitude} & \textbf{Filter} & \textbf{Telescope} & \textbf{Analysis} \\
(days) & ~ &  (AB) & ~ & ~ & ~\\
(1) & (2) & (3) & (4) & (5) & (6)\\
\hline 
0.00091 & 60613.24261 & 16.24 $\pm$ 0.27 & v & UVOT & X\\
0.00102 & 60613.242717 & 16.21 $\pm$ 0.04 & w & UVOT & X\\
0.00113 & 60613.242832 & 16.15 $\pm$ 0.03 & w & UVOT & X\\
0.00125 & 60613.242948 & 16.04 $\pm$ 0.03 & w & UVOT & X\\
0.00136 & 60613.243064 & 15.65 $\pm$ 0.09 & w & UVOT & X\\
0.00148 & 60613.243179 & 15.65 $\pm$ 0.09 & w & UVOT & X\\
0.00159 & 60613.243295 & 15.71 $\pm$ 0.08 & w & UVOT & X\\
0.00171 & 60613.243411 & 15.55 $\pm$ 0.09 & w & UVOT & X\\
0.00183 & 60613.243527 & 15.89 $\pm$ 0.02 & w & UVOT & X\\
0.00194 & 60613.243642 & 15.97 $\pm$ 0.02 & w & UVOT & X\\
0.00206 & 60613.243758 & 16.25 $\pm$ 0.04 & w & UVOT & X\\
0.00217 & 60613.243874 & 16.45 $\pm$ 0.04 & w & UVOT & X\\
0.00229 & 60613.24399 & 16.61 $\pm$ 0.05 & w & UVOT & X\\
0.0024 & 60613.244105 & 16.68 $\pm$ 0.05 & w & UVOT & X\\
0.00252 & 60613.244221 & 16.64 $\pm$ 0.05 & w & UVOT & X\\
0.00264 & 60613.244337 & 16.49 $\pm$ 0.04 & w & UVOT & X\\
0.00354 & 60613.245238 & 14.42 $\pm$ 0.03 & u & UVOT & X\\
0.00377 & 60613.245469 & 14.08 $\pm$ 0.03 & u & UVOT & X\\
0.004 & 60613.245701 & 13.77 $\pm$ 0.02 & u & UVOT & X\\
0.00423 & 60613.245932 & 13.66 $\pm$ 0.02 & u & UVOT & X\\
0.00446 & 60613.246164 & 13.65 $\pm$ 0.02 & u & UVOT & X\\
0.00469 & 60613.246395 & 13.62 $\pm$ 0.02 & u & UVOT & A-P\\
0.00493 & 60613.246627 & 13.79 $\pm$ 0.03 & u & UVOT & A-P\\
0.00516 & 60613.246858 & 13.88 $\pm$ 0.03 & u & UVOT & A-P\\
0.00539 & 60613.24709 & 13.9 $\pm$ 0.03 & u & UVOT & A-P\\
0.00562 & 60613.247321 & 14.0 $\pm$ 0.03 & u & UVOT & A-P\\
0.00585 & 60613.247553 & 14.05 $\pm$ 0.03 & u & UVOT & A-P\\
0.00608 & 60613.247784 & 14.05 $\pm$ 0.03 & u & UVOT & A-P\\
0.00631 & 60613.248016 & 14.2 $\pm$ 0.04 & u & UVOT & A-P\\
0.00649 & 60613.248188 & 13.72 $\pm$ 0.03 & b & UVOT & A-P\\
0.00669 & 60613.248388 & 14.43 $\pm$ 0.35 & w & UVOT & X\\
0.0068 & 60613.248504 & 14.42 $\pm$ 0.32 & w & UVOT & X\\
0.00692 & 60613.248619 & 14.5 $\pm$ 0.45 & w & UVOT & X\\
\bottomrule
\end{tabular}
\label{tab:op-1}
\end{table}

\setlength{\tabcolsep}{1pt}
\begin{table}[]
\centering
\caption{UV/Optical observations of GRB~241030A (continued from \ref{tab:op-1})} 
\begin{tabular}{cccccr}
\toprule
\hline
\boldmath $\Delta T$ \unboldmath & \textbf{MJD} & \textbf{Magnitude} & \textbf{Filter} & \textbf{Telescope} & \textbf{Analysis} \\
(days) & ~ &  (AB) & ~ & ~ & ~\\
(1) & (2) & (3) & (4) & (5) & (6)\\
\hline 
0.00713 & 60613.24883 & 16.89 $\pm$ 0.11 & uvw2 & UVOT & X\\
0.00716 & 60613.24886 & 13.52 $\pm$ 0.06 & v & UVOT & A-P\\
0.00751 & 60613.249207 & 13.66 $\pm$ 0.06 & v & UVOT & A-P\\
0.00787 & 60613.249571 & 16.31 $\pm$ 0.1 & uvm2 & UVOT & X\\
0.00798 & 60613.249683 & 15.59 $\pm$ 0.06 & uvw1 & UVOT & X\\
0.00817 & 60613.249867 & 14.58 $\pm$ 0.04 & u & UVOT & A-P\\
0.0084 & 60613.250099 & 14.52 $\pm$ 0.08 & u & UVOT & A-P\\
0.00847 & 60613.250167 & 14.06 $\pm$ 0.04 & b & UVOT & A-P\\
0.00865 & 60613.250355 & 14.82 $\pm$ 0.22 & w & UVOT & X\\
0.00869 & 60613.250387 & 14.21 $\pm$ 0.08 & b & UVOT & A-P\\
0.00877 & 60613.250471 & 14.84 $\pm$ 0.16 & w & UVOT & X\\
0.00886 & 60613.250566 & 17.52 $\pm$ 0.25 & uvw2 & UVOT & X\\
0.00889 & 60613.250587 & 14.84 $\pm$ 0.21 & w & UVOT & X\\
0.00921 & 60613.250913 & 17.43 $\pm$ 0.18 & uvw2 & UVOT & X\\
0.00924 & 60613.250943 & 14.04 $\pm$ 0.06 & v & UVOT & A-P\\
0.00926 & 60613.250959 & 16.82 $\pm$ 0.23 & uvm2 & UVOT & X\\
0.00959 & 60613.25129 & 13.83 $\pm$ 0.09 & v & UVOT & A-P\\
0.00972 & 60613.251419 & 16.15 $\pm$ 0.14 & uvw1 & UVOT & X\\
0.00995 & 60613.251654 & 16.6 $\pm$ 0.14 & uvm2 & UVOT & X\\
0.01007 & 60613.251766 & 15.86 $\pm$ 0.08 & uvw1 & UVOT & X\\
0.01859 & 60613.260289 & 14.96 $\pm$ 0.01 & R & TRT-SRO & A-P\\
0.01945 & 60613.261148 & 15.05 $\pm$ 0.01 & R & TRT-SRO & A-P\\
0.02029 & 60613.261996 & 15.08 $\pm$ 0.01 & R & TRT-SRO & A-P\\
0.02316 & 60613.26486 & 15.29 $\pm$ 0.01 & R & TRT-SRO & A-P\\
0.02485 & 60613.266556 & 15.4 $\pm$ 0.01 & R & TRT-SRO & A-P\\
0.02653 & 60613.268234 & 15.47 $\pm$ 0.01 & R & TRT-SRO & A-P\\
0.02822 & 60613.269922 & 15.56 $\pm$ 0.01 & R & TRT-SRO & A-P\\
0.03309 & 60613.274792 & 15.72 $\pm$ 0.01 & R & TRT-SRO & A-P\\
0.0356 & 60613.2773 & 15.8 $\pm$ 0.01 & R & TRT-SRO & A-P\\
0.03809 & 60613.279796 & 15.88 $\pm$ 0.01 & R & TRT-SRO & A-P\\
0.04061 & 60613.282308 & 15.96 $\pm$ 0.01 & R & TRT-SRO & A-P\\
0.0431 & 60613.284797 & 16.0 $\pm$ 0.01 & R & TRT-SRO & A-P\\
0.04558 & 60613.287284 & 16.08 $\pm$ 0.01 & R & TRT-SRO & A-P\\
0.04809 & 60613.289789 & 16.15 $\pm$ 0.01 & R & TRT-SRO & A-P\\
0.05095 & 60613.292653 & 16.21 $\pm$ 0.01 & R & TRT-SRO & A-P\\
\bottomrule
\end{tabular}
\label{tab:op-2}
\end{table}

\setlength{\tabcolsep}{1pt}
\begin{table}[]
\centering
\caption{UV/Optical observations of GRB~241030A (continued from \ref{tab:op-2})} 
\begin{tabular}{cccccr}
\toprule
\hline
\boldmath $\Delta T$ \unboldmath & \textbf{MJD} & \textbf{Magnitude} & \textbf{Filter} & \textbf{Telescope} & \textbf{Analysis} \\
(days) & ~ &  (AB) & ~ & ~ & ~\\
(1) & (2) & (3) & (4) & (5) & (6)\\
\hline 
0.05389 & 60613.295587 & 16.42 $\pm$ 0.05 & r' & COLIBRÍ & A-P\\
0.05451 & 60613.296211 & 16.31 $\pm$ 0.01 & R & TRT-SRO & A-P\\
0.05805 & 60613.299752 & 16.36 $\pm$ 0.02 & R & TRT-SRO & A-P\\
0.06161 & 60613.303312 & 16.42 $\pm$ 0.01 & R & TRT-SRO & A-P\\
0.06829 & 60613.309987 & 16.56 $\pm$ 0.02 & R & TRT-SRO & A-P\\
0.07496 & 60613.316663 & 16.67 $\pm$ 0.02 & R & TRT-SRO & A-P\\
0.0776 & 60613.319305 & 16.73 $\pm$ 0.05 & r' & COLIBRÍ & A-P\\
0.08109 & 60613.322792 & 16.79 $\pm$ 0.05 & r' & COLIBRÍ & A-P\\
0.08457 & 60613.326273 & 16.83 $\pm$ 0.05 & r' & COLIBRÍ & A-P\\
0.08472 & 60613.326419 & 16.81 $\pm$ 0.02 & R & TRT-SRO & A-P\\
0.08819 & 60613.329893 & 16.9 $\pm$ 0.05 & r' & COLIBRÍ & A-P\\
0.09139 & 60613.33309 & 16.89 $\pm$ 0.02 & R & TRT-SRO & A-P\\
0.09205 & 60613.333755 & 16.96 $\pm$ 0.05 & r' & COLIBRÍ & A-P\\
0.09585 & 60613.337554 & 16.98 $\pm$ 0.05 & r' & COLIBRÍ & A-P\\
0.09809 & 60613.33979 & 17.02 $\pm$ 0.02 & R & TRT-SRO & A-P\\
0.09978 & 60613.34148 & 17.06 $\pm$ 0.05 & r' & COLIBRÍ & A-P\\
0.10372 & 60613.345418 & 17.13 $\pm$ 0.05 & r' & COLIBRÍ & A-P\\
0.10478 & 60613.346482 & 17.09 $\pm$ 0.02 & R & TRT-SRO & A-P\\
0.10716 & 60613.348864 & 17.18 $\pm$ 0.05 & r' & COLIBRÍ & A-P\\
0.11147 & 60613.35317 & 17.16 $\pm$ 0.02 & R & TRT-SRO & A-P\\
0.11816 & 60613.359859 & 17.3 $\pm$ 0.04 & R & TRT-SRO & A-P\\
0.11915 & 60613.360847 & 18.14 $\pm$ 0.05 & u & UVOT & A-P\\
0.12266 & 60613.364361 & 18.19 $\pm$ 0.05 & u & UVOT & A-P\\
0.12482 & 60613.36652 & 17.35 $\pm$ 0.05 & R & TRT-SRO & A-P\\
0.12617 & 60613.367876 & 18.28 $\pm$ 0.05 & u & UVOT & A-P\\
0.12871 & 60613.370408 & 17.8 $\pm$ 0.08 & b & UVOT & A-P\\
0.1315 & 60613.373205 & 17.9 $\pm$ 0.05 & B & TRT-SRO & A-P\\
0.14236 & 60613.384065 & 17.71 $\pm$ 0.04 & V & TRT-SRO & A-P\\
0.14601 & 60613.387715 & 17.59 $\pm$ 0.04 & R & TRT-SRO & A-P\\
0.15327 & 60613.394973 & 17.16 $\pm$ 0.06 & I & TRT-SRO & A-P\\
0.25098 & 60613.492679 & 20.25 $\pm$ 0.17 & uvw1 & UVOT & X\\
0.3166 & 60613.558303 & 19.5 $\pm$ 0.07 & u & UVOT & A-P\\
0.32652 & 60613.568218 & 19.14 $\pm$ 0.09 & b & UVOT & A-P\\
0.37101 & 60613.612708 & 18.49 $\pm$ 0.06 & I & KNC & A-P\\
0.38187 & 60613.623569 & 21.61 $\pm$ 0.41 & uvm2 & UVOT & X\\
\bottomrule
\end{tabular}
\label{tab:op-3}
\end{table}

\setlength{\tabcolsep}{1pt}
\begin{table}[]
\centering
\caption{UV/Optical observations of GRB~241030A (continued from \ref{tab:op-3})} 
\begin{tabular}{cccccr}
\toprule
\hline
\boldmath $\Delta T$ \unboldmath & \textbf{MJD} & \textbf{Magnitude} & \textbf{Filter} & \textbf{Telescope} & \textbf{Analysis} \\
(days) & ~ &  (AB) & ~ & ~ & ~\\
(1) & (2) & (3) & (4) & (5) & (6)\\
\hline 
0.38192 & 60613.623623 & 18.92 $\pm$ 0.04 & R & AbAO-T70 & A-P\\
0.39121 & 60613.632913 & 20.74 $\pm$ 0.15 & uvw1 & UVOT & X\\
0.4485 & 60613.690206 & 20.14 $\pm$ 0.11 & u & UVOT & A-P\\
0.45857 & 60613.700272 & 20.13 $\pm$ 0.19 & b & UVOT & A-P\\
0.4998 & 60613.741503 & 19.06 $\pm$ 0.08 & r' & KAO & A-P\\
0.50522 & 60613.746921 & 18.94 $\pm$ 0.05 & i' & KAO & A-P\\
0.51064 & 60613.752342 & 18.75 $\pm$ 0.04 & z' & KAO & A-P\\
0.52452 & 60613.766222 & 21.04 $\pm$ 0.18 & uvw1 & UVOT & X\\
0.53737 & 60613.779069 & 19.65 $\pm$ 0.05 & g' & KNC & A-P\\
0.55271 & 60613.794411 & 19.31 $\pm$ 0.07 & r' & KNC & A-P\\
0.60878 & 60613.85048 & 19.68 $\pm$ 0.16 & r' & KNC & A-P\\
0.6637 & 60613.905404 & 19.9 $\pm$ 0.05 & g' & KNC & A-P\\
0.68166 & 60613.923361 & 19.75 $\pm$ 0.07 & R & HAO & A-P\\
0.68487 & 60613.92657 & 19.47 $\pm$ 0.06 & r' & KNC & A-P\\
0.70787 & 60613.949575 & 20.14 $\pm$ 0.07 & g' & KNC & A-P\\
0.7271 & 60613.968801 & 20.27 $\pm$ 0.06 & g' & HAO & A-P\\
0.72935 & 60613.97105 & 19.9 $\pm$ 0.1 & g' & KAO & A-P\\
0.7322 & 60613.9739 & 19.6 $\pm$ 0.13 & r' & KNC & A-P\\
0.74699 & 60613.988689 & 20.43 $\pm$ 0.2 & r' & KAO & A-P\\
0.75972 & 60614.001423 & 19.9 $\pm$ 0.07 & i' & KAO & A-P\\
0.77125 & 60614.012949 & 19.82 $\pm$ 0.08 & g' & KNC & A-P\\
0.77282 & 60614.014519 & 20.36 $\pm$ 0.07 & R & KNC & A-P\\
0.77584 & 60614.017546 & 21.55 $\pm$ 0.32 & uvw1 & UVOT & X\\
0.77628 & 60614.017986 & 19.76 $\pm$ 0.24 & R & \footnotesize{FR/CTAN} & A-P\\
0.78616 & 60614.027859 & 20.65 $\pm$ 0.16 & u & UVOT & A-P\\
0.79275 & 60614.034449 & 19.59 $\pm$ 0.14 & z' & KAO & A-P\\
0.8157 & 60614.057401 & 20.0 $\pm$ 0.25 & V & KNC & A-P\\
0.83943 & 60614.081134 & 19.95 $\pm$ 0.06 & r' & COLIBRÍ & A-P\\
0.84117 & 60614.082875 & 20.47 $\pm$ 0.28 & b & UVOT & A-P\\
0.85432 & 60614.096023 & 20.95 $\pm$ 0.3 & B & TRT-SRO & A-P\\
0.87975 & 60614.121453 & 19.71 $\pm$ 0.13 & V & TRT-SRO & A-P\\
0.90525 & 60614.146949 & 23.82 $\pm$ 2.11 & uvw2 & UVOT & X\\
0.91086 & 60614.152565 & 21.0 $\pm$ 2.94 & v & UVOT & A-P\\
0.9498 & 60614.191502 & 19.99 $\pm$ 0.06 & R & TRT-SRO & A-P\\
0.9643 & 60614.206005 & 19.67 $\pm$ 0.12 & I & TRT-SRO & A-P\\
\bottomrule
\end{tabular}
\label{tab:op-4}
\end{table}

\setlength{\tabcolsep}{1pt}
\begin{table}[]
\centering
\caption{UV/Optical observations of GRB~241030A (continued from \ref{tab:op-4})} 
\begin{tabular}{cccccr}
\toprule
\hline
\boldmath $\Delta T$ \unboldmath & \textbf{MJD} & \textbf{Magnitude} & \textbf{Filter} & \textbf{Telescope} & \textbf{Analysis} \\
(days) & ~ &  (AB) & ~ & ~ & ~\\
(1) & (2) & (3) & (4) & (5) & (6)\\
\hline 
0.96999 & 60614.211694 & 20.35 $\pm$ 0.31 & b & UVOT & A-P\\
1.13104 & 60614.372737 & 20.47 $\pm$ 0.23 & V & TRT-SRO & A-P\\
1.16712 & 60614.408819 & 22.26 $\pm$ 0.83 & uvm2 & UVOT & X\\
1.17731 & 60614.419011 & 23.63 $\pm$ 1.42 & uvw1 & UVOT & X\\
1.18869 & 60614.430387 & 20.62 $\pm$ 0.1 & r' & COLIBRÍ & A-P\\
1.2294 & 60614.4711 & 20.45 $\pm$ 0.06 & r' & COLIBRÍ & A-P\\
1.22986 & 60614.471558 & 21.04 $\pm$ 0.48 & u & UVOT & A-P\\
1.23027 & 60614.471968 & 20.66 $\pm$ 0.29 & r' & COLIBRÍ & A-P\\
1.23271 & 60614.474416 & 20.4 $\pm$ 0.22 & R & TRT-SRO & A-P\\
1.24359 & 60614.485294 & 19.7 $\pm$ 0.19 & I & TRT-SRO & A-P\\
1.2545 & 60614.496205 & 20.7 $\pm$ 0.2 & B & TRT-SRO & A-P\\
1.25696 & 60614.49866 & 20.7 $\pm$ 0.1 & r' & COLIBRÍ & A-P\\
1.40663 & 60614.648327 & 20.0 $\pm$ 0.1 & i' & TNOT & A-P\\
1.52624 & 60614.767941 & 20.34 $\pm$ 0.06 & i' & AbAO-T150 & X\\
1.52714 & 60614.768844 & 20.63 $\pm$ 0.05 & r' & AbAO-T150 & X\\
1.65471 & 60614.896412 & 20.8 $\pm$ 0.1 & i' & KAO & A-P\\
1.67051 & 60614.912209 & 20.99 $\pm$ 0.19 & R & HAO & A-P\\
1.6857 & 60614.927398 & 21.5 $\pm$ 0.1 & g' & KAO & A-P\\
1.72339 & 60614.965087 & 20.7 $\pm$ 0.16 & r' & KAO & A-P\\
1.86734 & 60615.109039 & 21.52 $\pm$ 0.2 & r' & COLIBRÍ & A-P\\
2.18275 & 60615.424456 & 20.57 $\pm$ 0.04 & r' & TNOT & A-P\\
2.49951 & 60615.741216 & 21.6 $\pm$ 0.2 & i' & KAO & A-P\\
2.52714 & 60615.768394 & 21.83 $\pm$ 0.1 & r' & AbAO-T150 & X\\
2.57294 & 60615.814643 & 21.43 $\pm$ 0.11 & r' & KAO & A-P\\
7.84252 & 60621.084225 & 22.62 $\pm$ 0.15 & R & NAO-2m & A-P\\
8.46644 & 60621.708137 & 23.86 $\pm$ 0.28 & R & NAO-1.5m & A-P\\
\bottomrule
\end{tabular}
\label{tab:op-5}
\end{table}

\end{document}